\documentstyle[12pt,epsf]{article}
\catcode`\@=11

\def\section{\@startsection{section}{1}{\z@}
  {3.5ex plus 1.0ex minus 0.2ex}{2.3ex plus .2ex}{\normalsize}}
\def\subsection{\@startsection{subsection}{2}{\z@}
  {3.25ex plus 1.0ex minus 0.2ex}{1.5ex plus 0.2ex}{\normalsize\bf}}
\def\subsubsection{\@startsection{subsubsection}{3}{\z@}
  {3.25ex plus 1.0ex minus 0.2ex}{1.5ex plus 0.2ex}{\normalsize\bf}}

\@addtoreset{equation}{section}

\catcode`\@=12

\newcommand{\skipline}{\vspace{\baselineskip}}

\def\fun#1#2{\lower3.6pt\vbox{\baselineskip0pt\lineskip.9pt
  \ialign{$\mathsurround=0pt#1\hfil##\hfil$\crcr#2\crcr\sim\crcr}}}

\begin{document}

\begin{titlepage}

\begin{center}
Meson Decay Constants from the Valence Approximation\\
to Lattice QCD
\end{center}

\skipline
\skipline

\begin{center}

F. Butler, H. Chen, 
J. Sexton\footnote{permanent address: Department of Mathematics,
Trinity College, Dublin 2,
Republic of Ireland}, 
A. Vaccarino,\\
and D. Weingarten \\

IBM Research \\
P.O. Box 218, Yorktown Heights, NY 10598
\end{center}

\skipline
\skipline

\begin{center}
ABSTRACT
\end{center}

\begin{quotation}

We evaluate $f_{\pi}/ m_{\rho}$, $f_K/ m_{\rho}$, $1/f_{\rho}$, and $
m_{\phi}/(f_{\phi} m_{\rho})$, extrapolated to physical quark mass, zero
lattice spacing and infinite volume, for lattice QCD with Wilson quarks
in the valence (quenched) approximation.  The predicted ratios differ
from experiment by amounts ranging from 12\% to 17\% equivalent to
between 0.9 and 2.8 times the corresponding statistical uncertainties.

\end{quotation}

\skipline
\skipline

\end{titlepage}

\section{INTRODUCTION}

In a recent paper~\cite{Butler93} we presented lattice QCD predictions
for the masses of eight low-lying hardrons, extrapolated to physical
quark mass, zero lattice spacing, and infinite volume, using Wilson
quarks in the valence (quenched) approximation. The masses we found were
within 6\% of experiment, and all differences between prediction and
experiment were consistent with the calculation's statistical
uncertainty.  We argued that this result could be interpreted as
quantitative confirmation of the low-lying mass predictions both of QCD
and of the valence approximation. It appeared to us unlikely that eight
different valence approximation masses would agree with experiment yet
differ significantly from QCD's predictions including the full effect of
quark-antiquark vacuum polarization.

We have now evaluated the infinite volume, zero lattice spacing,
physical quark mass limit of $f_{\pi}/ m_{\rho}$, $f_K/ m_{\rho}$,
$1/f_{\rho}$, and $m_{\phi}/(f_{\phi} m_{\rho})$.  To our knowledge
there have been no previous systematic calculations of this physical
limit of lattice meson decay constants. A review of earlier work is
given in Ref. \cite{Toussaint}.  The predicted $m_{\phi}/(f_{\phi}
m_{\rho})$ lies above its observed value by about its statistical
uncertainty of approximately 15\%.  The predicted $f_{\pi}/ m_{\rho}$
and $1/f_{\rho}$ are close to 12\% below experiment, equivalent to about
0.9 times the corresponding statistical uncertainties, and $f_K/
m_{\rho}$ is below experiment by 17\%, equal to 2.8 times its
statistical uncertainty.

Although overall the predicted values could be considered in fair
agreement with experiment, the result that three of four decay constants
range from 0.9 to 2.8 standard deviations below experiment suggests the
possiblity that the true, underlying values, if determined without
statistical uncertainties, would fall somewhat below experiment.  One
possible source of such disagreement is that we have assumed exact
isospin symmetry and ignored electromagnetic effects. In particular,
although our prediction for $f_{\pi}$ lies below measured values
determined from charged pion decays, reported numbers for the neutral
pion decay constant \cite{Cello} are quite close to our result.  The
significance of the closer agreement between our predicted $f_{\pi}$ and
the neutral pion decay constant is unclear to us, however, as a
consequence of the systematic uncertainties arising in the experimental
determination of the neutral pion decay constant.

Another possible source of disagreement between our numbers and
experiment may be the valence approximation itself.  A simple physical
argument tends to support of this alternative
\cite{Fermilab}.  The valence approximation may be viewed as replacing
the momentum and frequency dependent color dielectric constant arising
from quark-antiquark vacuum polarization with its low momentum limit
\cite{Weingar81}.  At low momentum, then, the effective quark charge
appearing in the valence approximation will agree with the low momentum
effective charge of the full theory.  The valence approximation might
thus be expected to be fairly reliable for low-lying baryon and meson
masses, which are determined largely by the long distance behavior of
the chromoelectric field.  The valence approximation's effective quark
charge at higher momentum can be obtained from the low momentum charge
by the Callan-Symanzik equation.  As a consequence of the absence of
dynamical quark-antiquark vacuum polarization, the quark charge in the
valence approximation will fall faster with momentum than it does in the
full theory, at short distance the attractive quark-antiquark potential
in the valence approximation will be weaker than in the full theory, and
meson wave functions in the valence approximation will be pulled into
the origin less than in the full theory.  Since decay constants are
proportional to the square of wave functions at the origin, decay
constants in the valence approximation could be expected to be smaller
than in the full theory.

The calculations described here were done on the GF11 parallel computer
at IBM Research \cite{Weingar90} and use the same collection of gauge
configurations and quark propagators generated for the mass calculations
of Ref. \cite{Butler93}.  The full set of mass and decay constant
calculations took approximately one year to complete.  GF11 was used in
configurations ranging from 384 to 480 processors, with sustained speeds
ranging from 5 Gflops to 7 Gflops.  With the present set of improved
algorithms and 480 processors, these calculations could be repeated in
less than four months.

\section{DEFINITIONS}\label{sect:defs}

The normalization we adopt for pseudoscalar and vector decay constants
in continuum QCD is 
\begin{eqnarray}
<\! 0 | J^{\mu}_j(0) | V(p, \epsilon, j)\! > \; & = & \epsilon^{\mu} m_j
F_j, \nonumber \\
<\! 0 | J^{5 \mu}_j(0) | P(p, j)\! > & = & p^{\mu} f_j, \nonumber
\end{eqnarray}
for vector and pseudoscalar states, $| V(p, \epsilon, j)\! >$ and $|
P(p, j)\! >$, respectively, normalized by
\begin{eqnarray}
< \!p | q \!> \; = (2 \pi)^3 p^0 \delta( \vec{p} - \vec{q}). \nonumber
\end{eqnarray}
Here $j$ is a flavor-SU(3) octet index and vector and axial vector
flavor-SU(3) currents $J^{\mu}_j(x)$ and $J^{5 \mu}_j(x)$ are related to
cannonical continuum quark and antiquark fields, $\overline{\psi}_c$
and ${\psi}_c$, and an orthonormal set of
flavor-SU(3) matrices $\lambda_j$ by
\begin{eqnarray}
J^{\mu}_j(x) & = & \overline{\psi}_c(x) \gamma^{\mu} \lambda_j
\psi_c(x), \nonumber \\ J^{5 \mu}_j(x) & = & \overline{\psi}_c(x)
\gamma^5 \gamma^{\mu} \lambda_j
\psi_c(x), \nonumber \\
tr( \lambda_j \lambda_j) & = & \frac{1}{2}. \nonumber
\end{eqnarray}
Assuming exact isospin symmetry, we have 
\begin{center}
\begin{tabular}{ll}
$f_i =  f_{\pi}$ & $i = 1, \ldots 3,$ \\
$f_i = f_K$ & $i = 4, \ldots 7,$ \\
$F_i  = \frac{m_{\rho}}{ f_{\rho}}$ & $i = 1, \ldots 3.$  
\end{tabular}
\end{center}
In the valence approximation we have, in addition,
\begin{eqnarray}
F_8 = \frac{3 m_{\phi}}{\sqrt{2} f_{\phi}} \nonumber.
\end{eqnarray}
For simplicity, we will also
use the names $F_{\rho}$ and $F_{\phi}$ for $F_1$ and $F_8$,
respectively. Our normalization gives $f_{\pi}$ the value $(93.15 \pm
0.11)$ MeV.

The hadron mass calculation of Ref. \cite{Butler93} was done using
gaussian smeared lattice quark and antiquark fields defined in Coulomb gauge.
Smeared fields have, therefore, also been adopted for the present
calculation.  The smeared field $\phi_r(\vec{x},t)$ is related by
the lattice antiquark field $\psi_{\ell}(\vec{x},t)$ by
\begin{eqnarray}
\phi_r(\vec{x},t) & = & \sum_{\vec{y}} G_r(\vec{x} - \vec{y})
\psi_{\ell}(\vec{y},t), \nonumber \\ 
G_r(\vec{z}) & = & (\sqrt{\pi}r)^{-3} exp( - \frac{|\vec{z}|^2}{ r^2}).
\nonumber 
\end{eqnarray}
The field $\overline{\phi}_r(\vec{x},t)$ is defined correspondingly
from $\overline{\psi}_{\ell}(\vec{x},t)$.  We
take the smeared fields $\phi_0(x)$ and $\overline{\phi}_0(x)$ to be
$\psi(x)$ and $\overline{\psi}(x)$, respectively. From these fields,
define smeared currents by
\begin{eqnarray}
J^5_{j r}(x) & = & \overline{\phi}_r(x) \gamma^5 \lambda_j \phi_r(x), \nonumber \\
J^{\mu}_{j r}(x) & = & \overline{\phi}_r(x) \gamma^{\mu} \lambda_j \phi_r(x) ,
\nonumber \\
J^{5 \mu}_{j r}(x) & = & \overline{\phi}_r(x) \gamma^5 \gamma^{\mu} \lambda_j
\phi_r(x) . \nonumber
\end{eqnarray}

Define the correlation functions
$C^P_{j r' r}(t)$, $C^V_{j r' r}(t)$  and
$C^A_{j r' r}(t)$ to be
\begin{eqnarray}
C^P_{j r' r}(t) & = &
\sum_{\vec{x}} <\! [J^5_{ j r'}(\vec{x},t)]^{\dagger} J^5_{j
r}(0,0)\!>, \\
C^V_{j r' r}(t) & = &
\sum_{\vec{x}} <\![J^i_{j r'}(\vec{x},t)]^{\dagger} J^i_{ j r}(0,0)\!>, \\
C^A_{j r' r}(t) & = &
\sum_{\vec{x}} <\![J^{5 0}_{ j r'}(\vec{x},t)]^{\dagger} J^{5 0}_{j r}(0,0)\!>.
\end{eqnarray}
Then for a set of constants $Z^P_{j r' r}$, $Z^V_{j r' r}$  and
$Z^A_{j r' r}$, these correlation functions
have the asymptotic forms, for a large time separation
$t$ and lattice time direction period $T$,
\begin{eqnarray}
\label{asymP}
C^P_{j r' r}(t) 
& \rightarrow & Z^P_{j r' r} \{ exp( -m^P_j t) + exp[ -m^P_j (T - t)]\}, \\
\label{asymV}
C^V_{j r' r}(t)
& \rightarrow & Z^V_{j r' r}\{ exp( -m^V_j t) + exp[ -m^V_j (T -
t)]\} \\
\label{asymA}
C^A_{j r' r}(t)
& \rightarrow & Z^A_{j r' r} \{ exp( -m^P_j t) + exp[ -m^P_j (T - t)]\}. 
\end{eqnarray}
Here $m^P_j$ and $m^V_j$ are pseudoscalar and
vector masses, respectively. Eqs. (\ref{asymP}) and (\ref{asymA}) imply, in addition,
the asymptotic form
\begin{eqnarray}
\label{asymAoverP}
\frac{C^A_{j r' r''}(t)}{C^P_{j r' r}(t)} \rightarrow 
\frac{Z^A_{j r' r''}}{Z^P_{j r' r}}.
\end{eqnarray}

Measured in units of the lattice spacing $a$ the decay constants $f_j a$
and $F_j a$ are then given, for any choice of smearing size $r$, by
\begin{eqnarray}
\label{deffj}
(f_j a)^2 & = &
\frac{ 2 (z^A_j Z^{AP}_{j 0 r})^2} 
{m_j a Z^P_{jrr}} \\
\label{defFj}
(F_j a)^2 & = &
\frac{ 2 (z^V_j Z^V_{j 0 r})^2}
{m_j a Z^V_{jrr}}.
\end{eqnarray}
The coefficients
$z^A_j$ and $z^V_j$ are finite renormalizations chosen so that the
lattice currents $z^A_j a^3 J^{5 \mu}_{j 0}$ and $z^V_j a^3 J^{\mu}_{j 0}$
approach the continuum currents $J^{5 \mu}_j$ and $J^{\mu}_j$,
respectively, as the lattice spacing approaches zero.  
These constants are often given the ``naive'' values
\begin{eqnarray}
\label{defnaive}
z^{A N}_j & = & 2 k^P_j, \nonumber \\
z^{V N}_j & = & 2 k^V_j. 
\end{eqnarray}
where $k^P_j$ and $k^V_j$ are the hopping constants corresponding to the
mass of the quark and antiquark for a pseudoscalar or vector meson,
respectively, with flavor $j$, assumed here to have $m_q =
m_{\overline{q}}$.  A standard derivation of the naive finite
renormalization constants, however, which relates the quark terms in the lattice
action to those in the continuum action is actually not correct \cite{Lepage}.
Altough the naive finite renormalization constants do lead to the correct continuum
limit, naive renormalization contributes to decay constant 
an additional extraneous
dependence on lattice spacing.
Numerical evidence for this behavior will be given in Section \ref{sect:contlim}.
A consistent 
calculation of finite renormalizations \cite{Lepage}, correct to zeroth
in a mean-field theory improved perturbation expansion, gives
\begin{eqnarray}
\label{defz0}
z^{A 0}_j & = & 1 - \frac{3 k^P_j}{4 k_c}, \nonumber \\
z^{V 0}_j & = & 1 - \frac{3 k^V_j}{4 k_c}, 
\end{eqnarray}
where $k_c$ is
the critical hopping constant at which the pion's mass becomes zero.
To first order in improved perturbation theory, the finite
renormalizations become $z^{A 1}_j z^{A 0}$ and
$z^{V 1}_j z^{V 0}$ where
\begin{eqnarray}
\label{defz1}
z^{A 1} & = & 1 - 0.31 \alpha_{\overline{ms}}( 1 / a),
\nonumber \\
z^{V 1} & = & 1 - 0.82 \alpha_{\overline{ms}}( 1 / a).
\end{eqnarray}
Decay constants for mesons with $m_q \neq m_{\overline{q}}$ will be
discussed below.  

For equal values of the up and down quark masses,
$m_u$ and $m_d$, we adopt the convention that a missing flavor index
j always has the value 1. Thus
\begin{eqnarray}
C^P_{r' r}(t)  & = & C^P_{1 r' r}(t),  \\ \nonumber
Z^P_{r' r}     & = & Z^P_{1 r' r},     \\ \nonumber
C^V_{r' r}(t)  & = & C^V_{1 r' r}(t),  \\ \nonumber      
Z^V_{r' r}     & = & Z^V_{1 r' r},     \\ \nonumber
C^A_{r' r}(t)  & = & C^A_{1 r' r}(t),  \\ \nonumber      
Z^A_{r' r}     & = & Z^A_{1 r' r}.     \\ \nonumber
\end{eqnarray}

\section{PROPAGATORS}\label{sect:props}

Table~\ref{tab:lattices} lists the lattice sizes, parameter values,
sweeps skipped between gauge configurations, and number of
configurations used in the ensembles from which decay constants were
calculated. Gauge configurations were generated with the
Cabbibo-Marinari-Okawa algorithm. A variety of different test support
the expectation that the large number of sweeps run between successive
configurations were more than sufficient to produce statistically
independent values for all of the quantities required by the
present calculations.  A discussion of the algorithms by which quark
propagators were found is given in Ref. \cite{Butler93}.

Vector and pseudoscalar masses and the coefficients $Z^P_{r' r}$,
$Z^V_{r' r}$, and $Z^A_{r' r}$ were then determined by fitting
$C^P_{r' r}(t)$, $C^V_{r' r}(t)$, and $C^A_{r' r}(t)$ to
the asymptotic forms of Eqs. (\ref{asymP}) - (\ref{asymA}). As an
alternative means of determining $Z^A_{r' r}$ given a value of
$Z^P_{r' r}$, fits of $C^A_{r' r}(t) / C^P_{r' r}(t)$
were also made to the asymptotic form of Eqs. (\ref{asymAoverP}).  

\begin{figure}
\epsfxsize=\textwidth \epsfbox{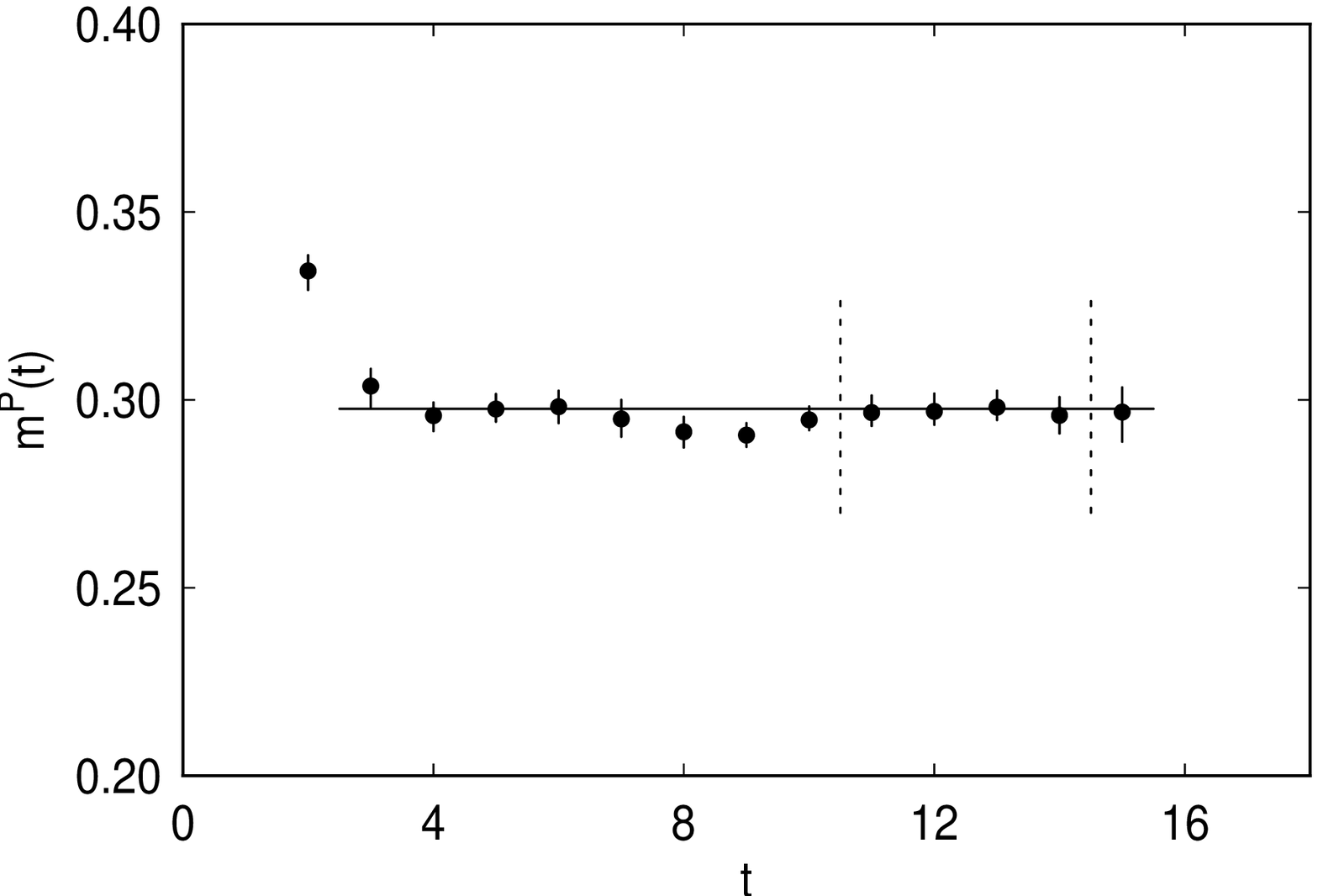}
\caption{ Effective masses and fitted mass for the pseudscalar propagator
$C^P_{0 2}(t)$ on the lattice $16^3 \times 32$ at 
at $\beta = 5.70$ and $k = 0.1675$
\label{fig:570mP}}
\end{figure}

\begin{figure}
\epsfxsize=\textwidth \epsfbox{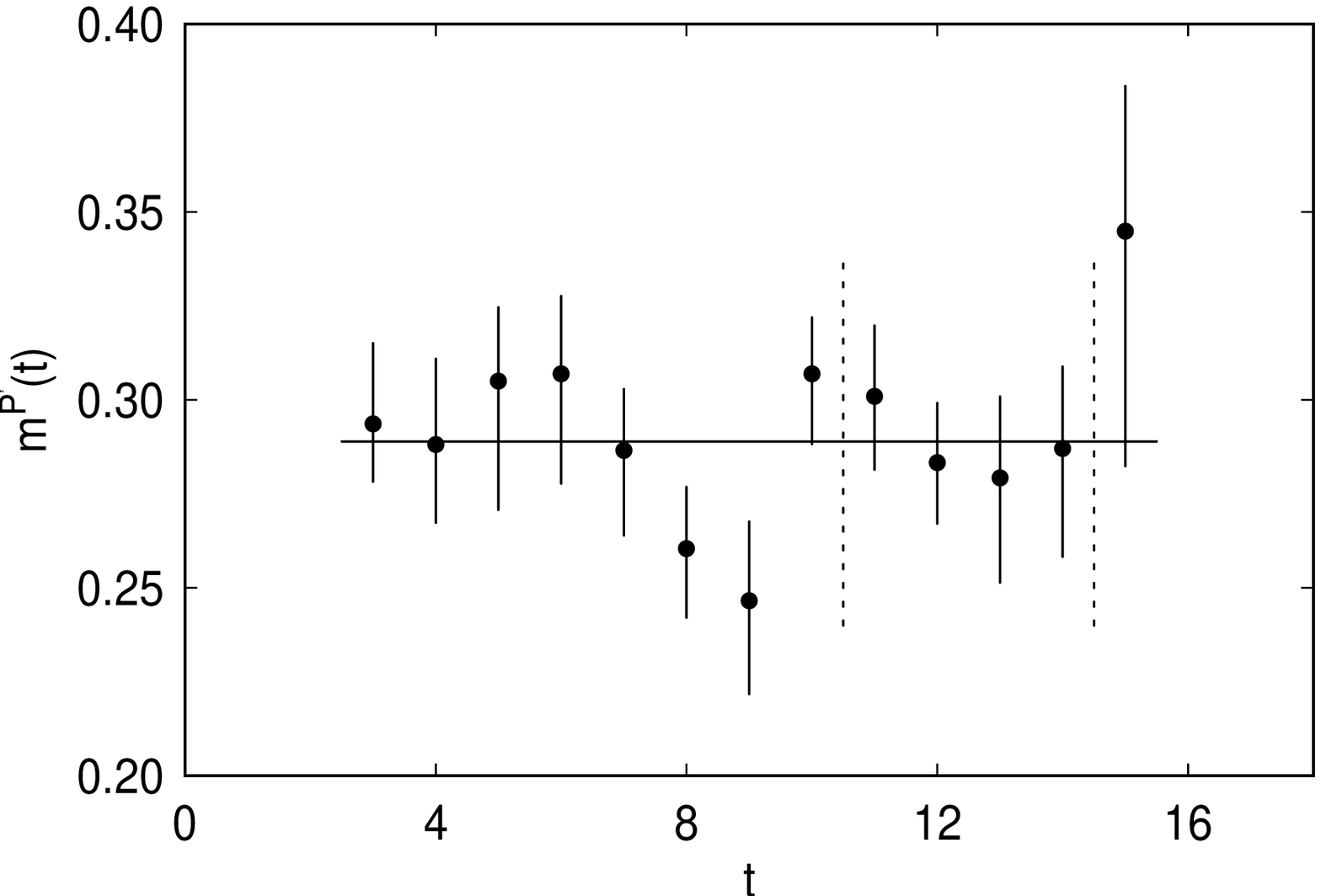}
\caption{ Effective masses and fitted mass for the axial vector propagator
$C^A_{0 2}(t)$ on the lattice $16^3 \times 32$ at 
at $\beta = 5.70$ and $k = 0.1675$
\label{fig:570mA}}
\end{figure}

\begin{figure}
\epsfxsize=\textwidth \epsfbox{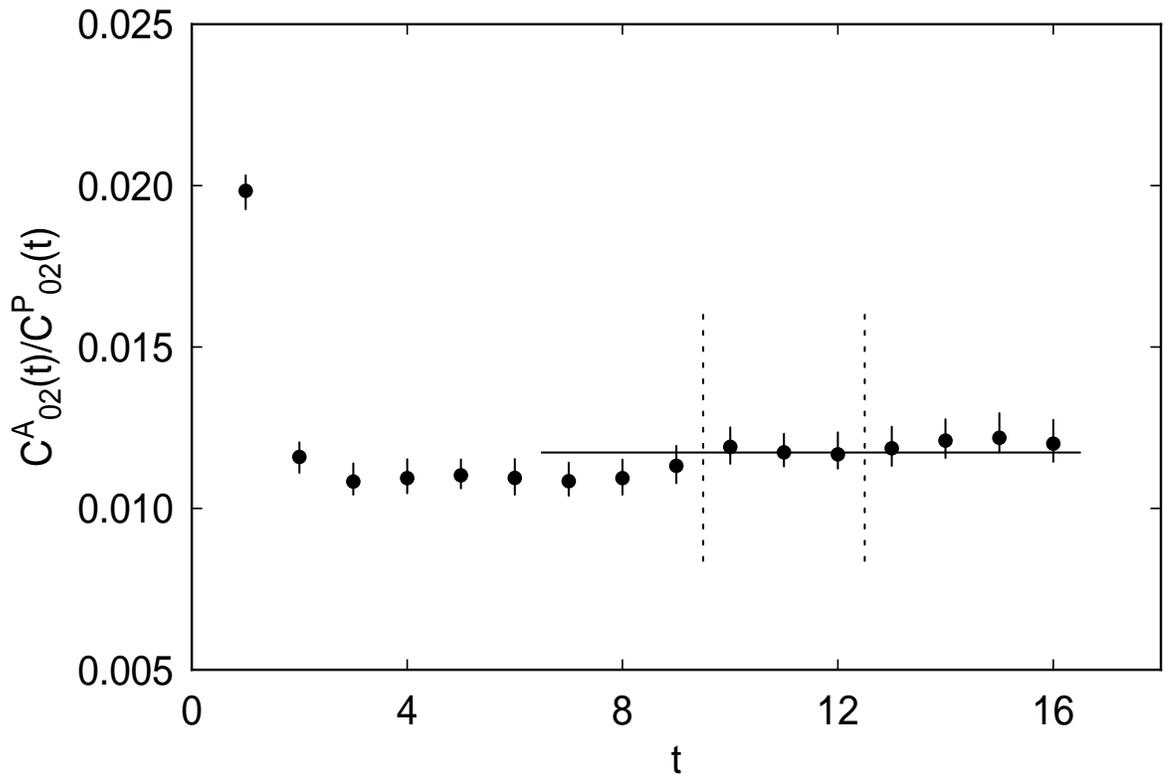}
\caption{ The propagator ratio $C^A_{0 2}(t) / C^P_{0 2}(t)$ 
and a fit to its large $t$ plateau on the lattice $16^3 \times 32$ 
at at $\beta = 5.70$ and $k = 0.1675$
\label{fig:570AP}}
\end{figure}

\begin{figure}
\epsfxsize=\textwidth \epsfbox{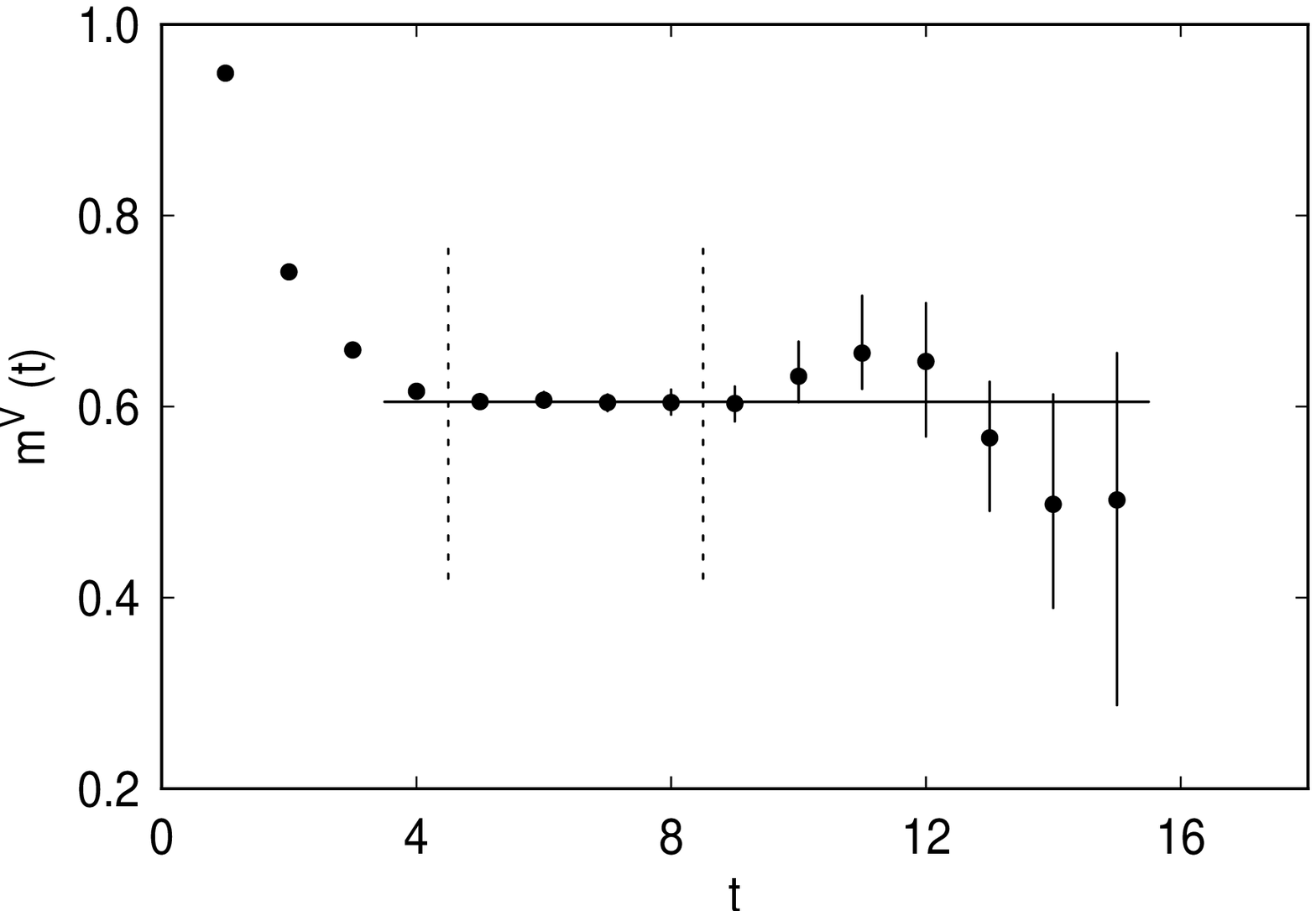}
\caption{ Effective masses and fitted mass for the vector propagator
$C^V_{0 2}(t)$ on the lattice $16^3 \times 32$ at 
at $\beta = 5.70$ and $k = 0.1675$
\label{fig:570mV}}
\end{figure}

\begin{figure}
\epsfxsize=\textwidth \epsfbox{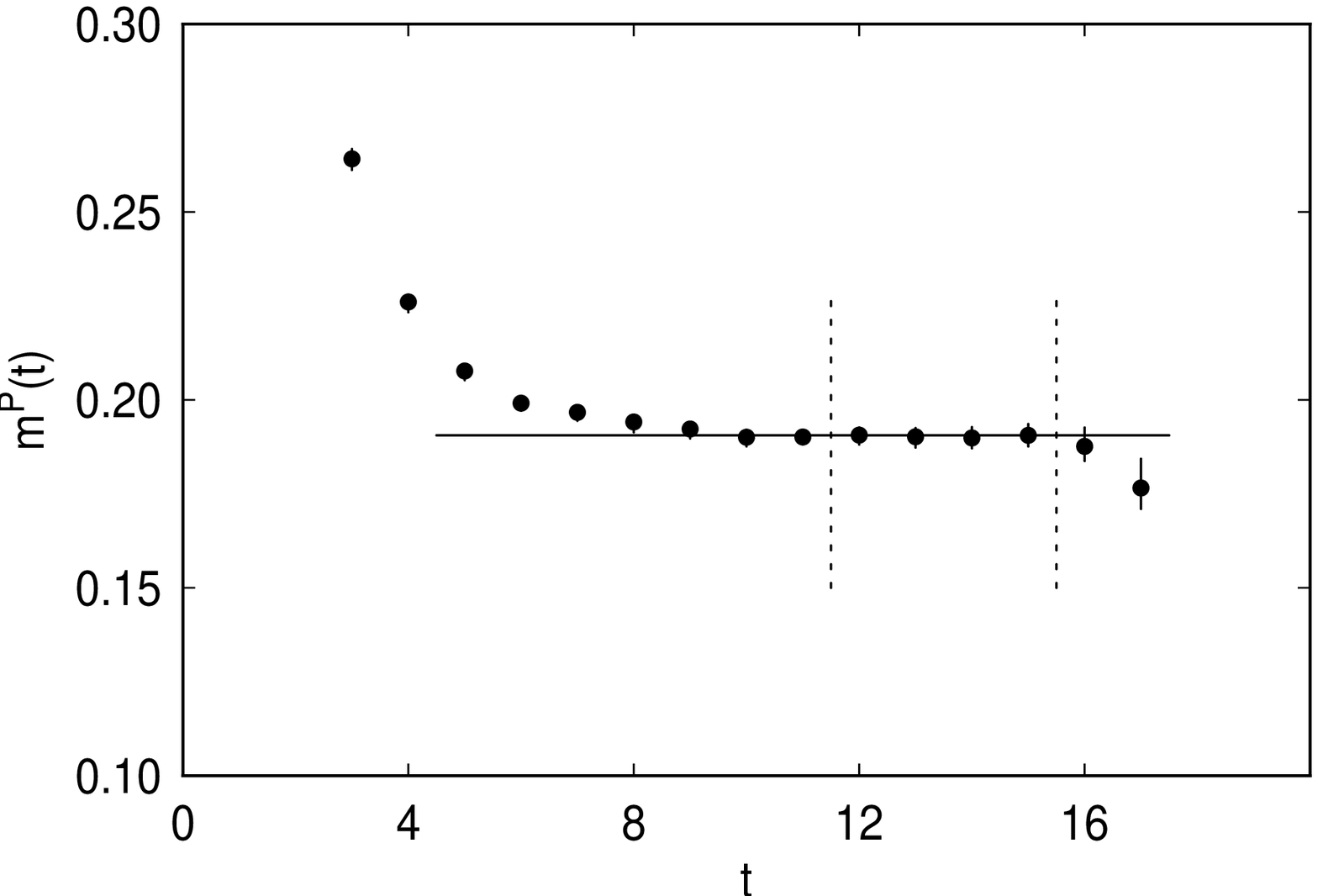}
\caption{ Effective masses and fitted mass for the pseudscalar propagator
$C^P_{0 2}(t)$ on the lattice $24^3 \times 36$ at 
at $\beta = 5.93$ and $k = 0.1581$
\label{fig:593mP}}
\end{figure}

\begin{figure}
\epsfxsize=\textwidth \epsfbox{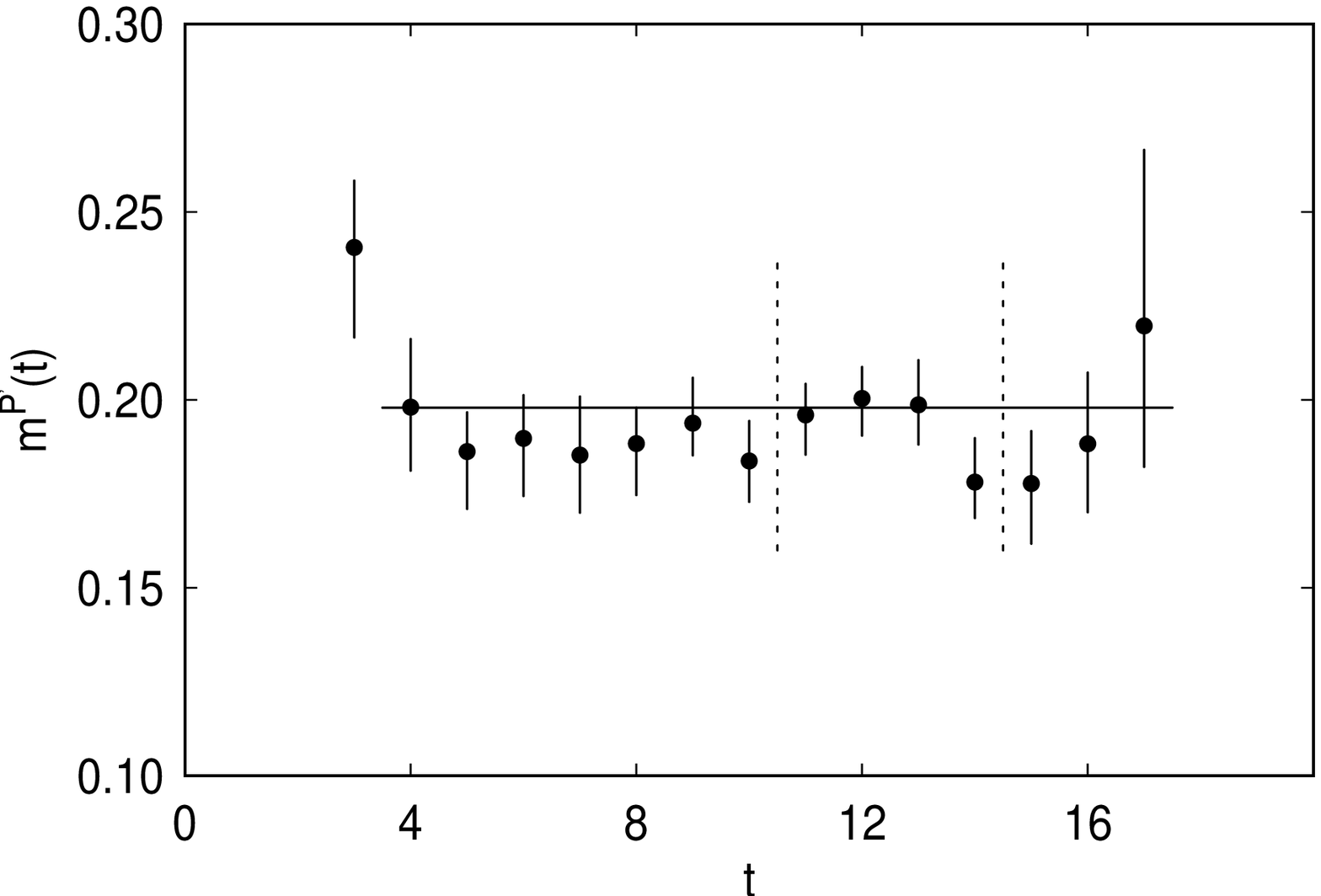}
\caption{ Effective masses and fitted mass for the axial vector propagator
$C^A_{0 2}(t)$ on the lattice $24^3 \times 36$ at 
at $\beta = 5.93$ and $k = 0.1581$
\label{fig:593mA}}
\end{figure}

\begin{figure}
\epsfxsize=\textwidth \epsfbox{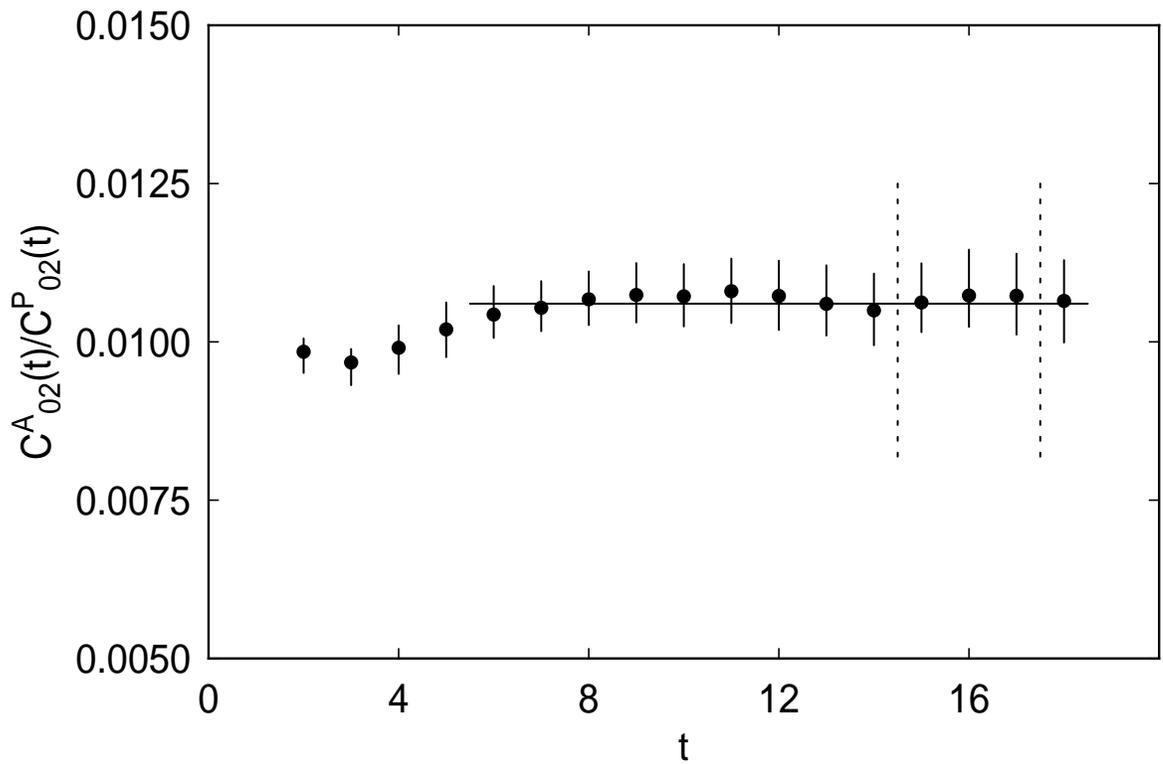}
\caption{ The propagator ratio $C^A_{0 2}(t) / C^P_{0 2}(t)$ 
and a fit to its large $t$ plateau on the lattice $24^3 \times 36$
at at $\beta = 5.93$ and $k = 0.1581$
\label{fig:593AP}}
\end{figure}

\begin{figure}
\epsfxsize=\textwidth \epsfbox{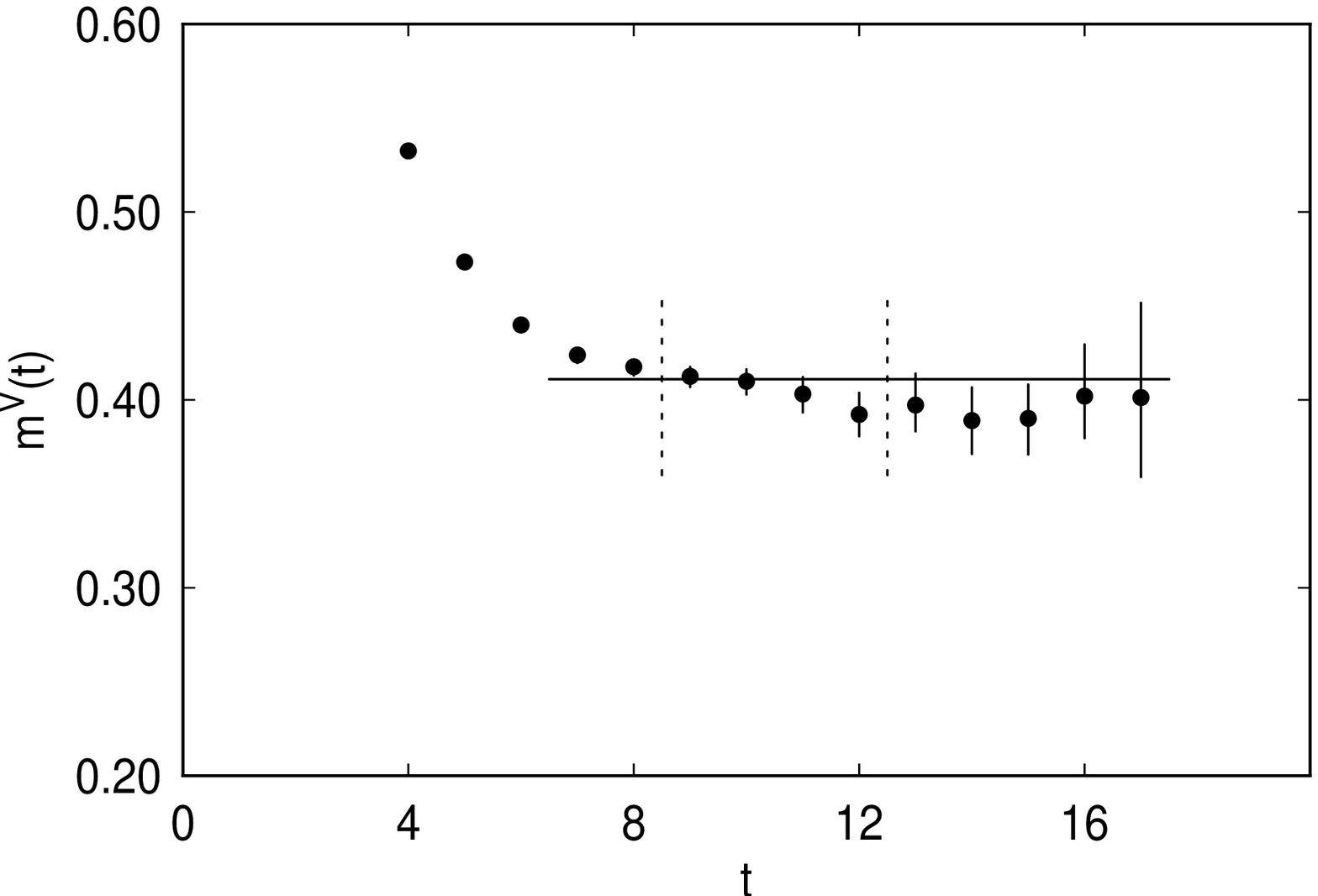}
\caption{ Effective masses and fitted mass for the vector propagator
$C^V_{0 2}(t)$ on the lattice $24^3 \times 36$ at 
at $\beta = 5.93$ and $k = 0.1581$
\label{fig:593mV}}
\end{figure}

\begin{figure}
\epsfxsize=\textwidth \epsfbox{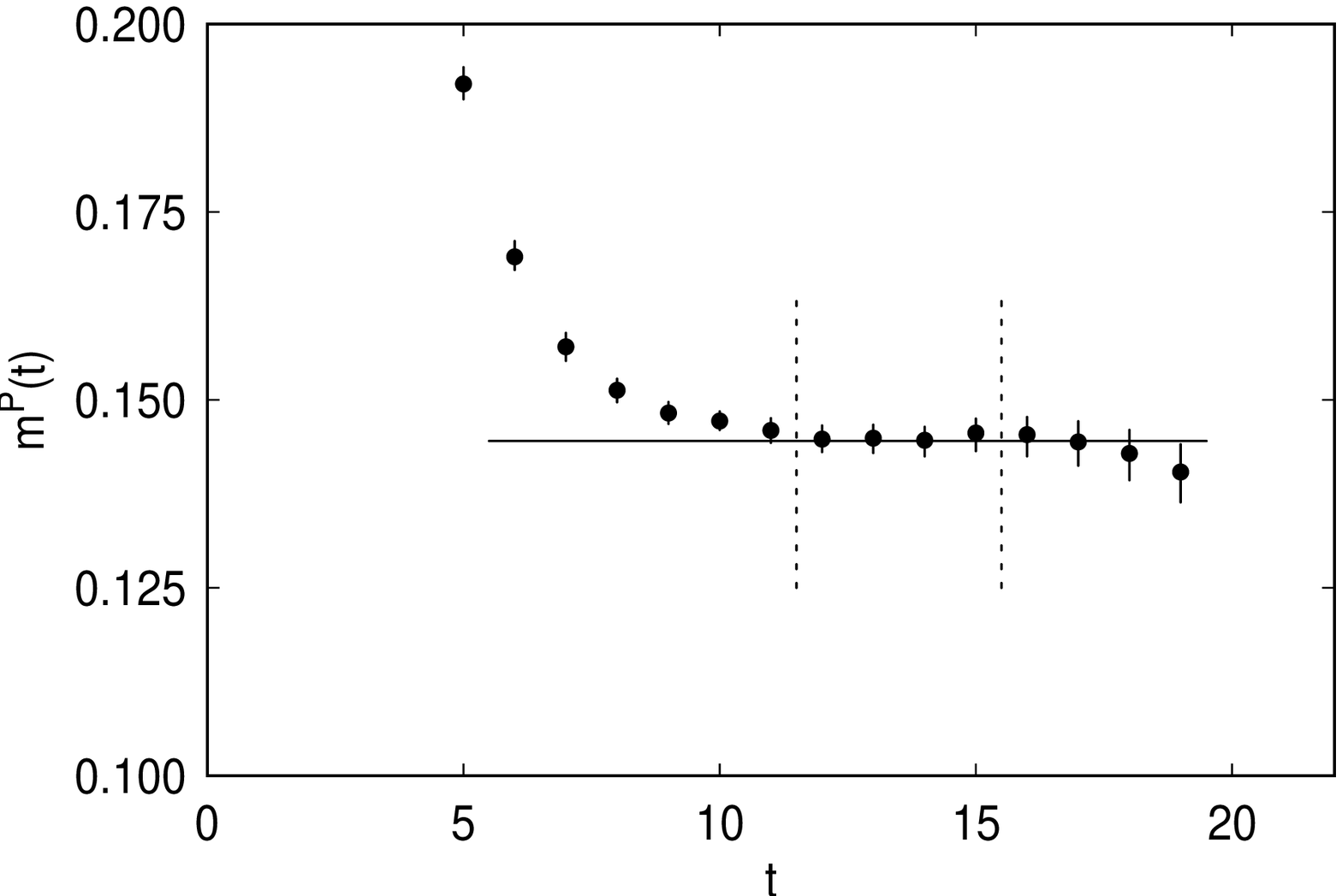}
\caption{ Effective masses and fitted mass for the pseudscalar propagator
$C^P_{0 2}(t)$ on the lattice $30 \times 32^2 \times 40$ at 
at $\beta = 6.17$ and $k = 0.1532$
\label{fig:617mP}}
\end{figure}

\begin{figure}
\epsfxsize=\textwidth \epsfbox{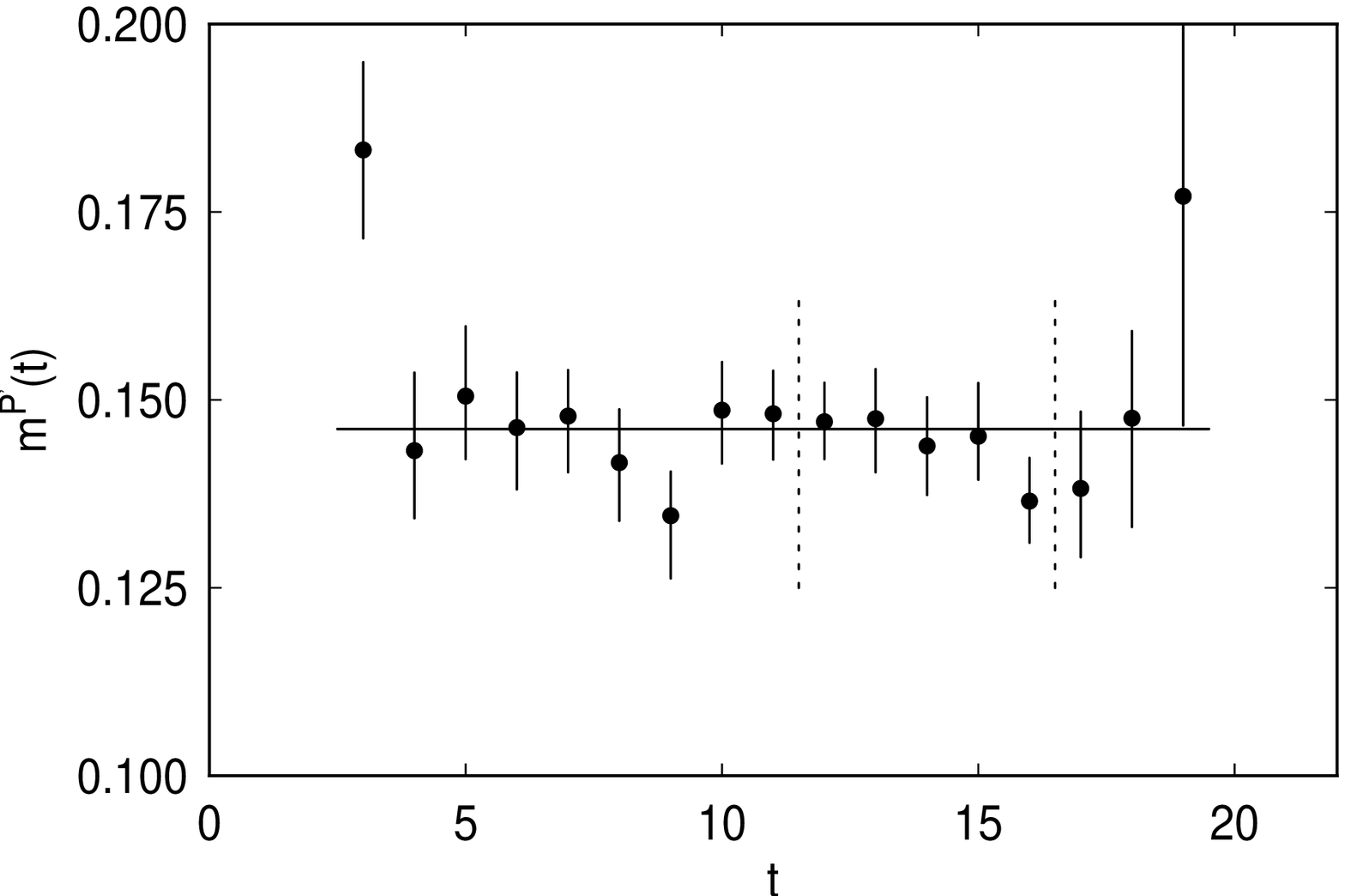}
\caption{ Effective masses and fitted mass for the axial vector propagator
$C^A_{0 2}(t)$ on the lattice $30 \times 32^2 \times 40$ at 
at $\beta = 6.17$ and $k = 0.1532$
\label{fig:617mA}}
\end{figure}

\begin{figure}
\epsfxsize=\textwidth \epsfbox{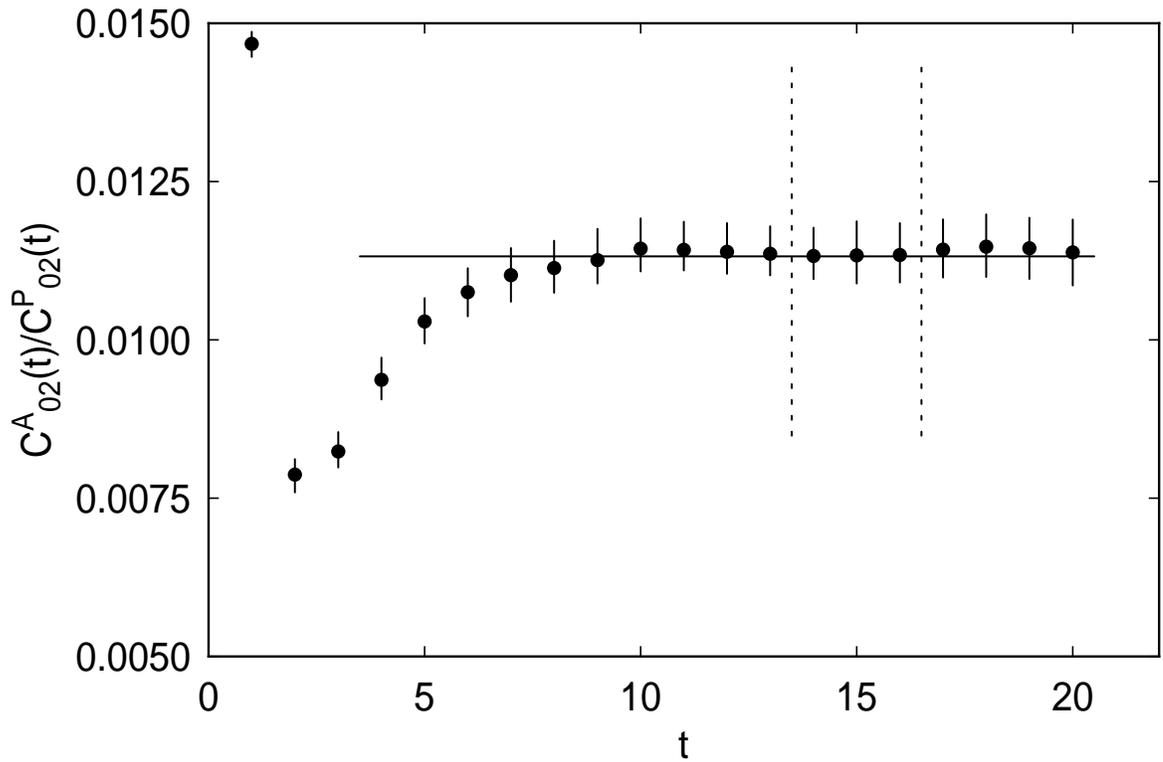}
\caption{ The propagator ratio $C^A_{0 2}(t) / C^P_{0 2}(t)$ 
and a fit to its large $t$ plateau on the lattice $30 \times 32^2 \times
40$ at at $\beta = 6.17$ and $k = 0.1532$
\label{fig:617AP}}
\end{figure}

\begin{figure}
\epsfxsize=\textwidth \epsfbox{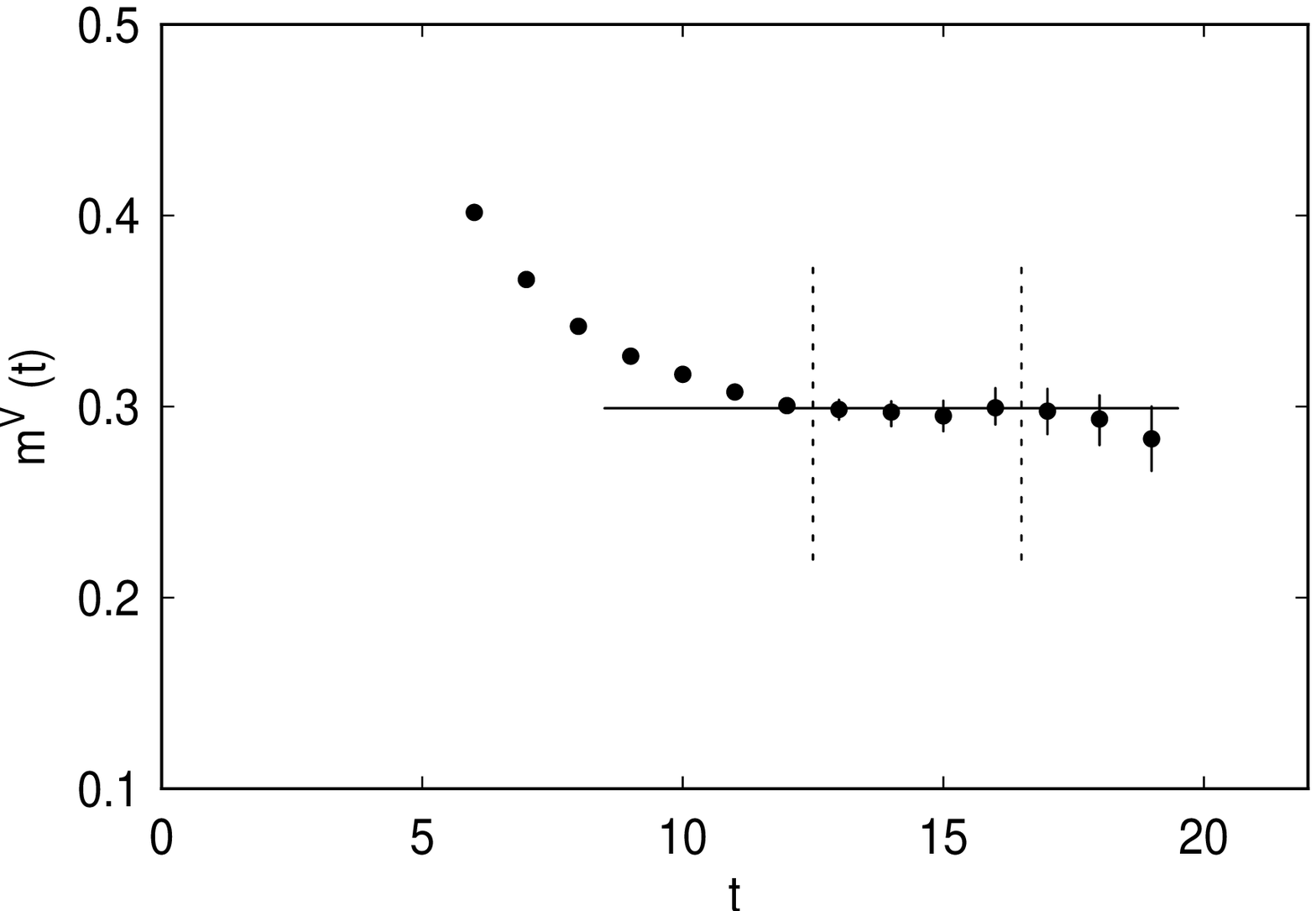}
\caption{ Effective masses and fitted mass for the vector propagator
$C^V_{0 2}(t)$ on the lattice $30 \times 32^2 \times 40$ at 
at $\beta = 6.17$ and $k = 0.1532$
\label{fig:617mV}}
\end{figure}

For the lattice $8^3 \times 32$ at $\beta$ of 5.70 we calculated
propagators only for source $r$ and sink $r'$ of size 0. In all other
cases we calculated propagators only for source size $r$ of 2. To
determine decay constants, according to Eqs. (\ref{deffj}) and
(\ref{defFj}), fits for the lattice $8^3 \times 32$ to propagators for
the single sink of size 0 are sufficient, while for all other lattices
fits are needed for sink sizes of both 0 and 2. To determine the range
of time separations to be used in fitting for each $\beta$ and $k$
value, we evaluated effective masses $m^P(t)$, $m^V(t)$ and $m^{P'}(t)$
by fitting $C^P_{ r' r}(t)$, $C^V_{ r' r}(t)$, and $C^A_{ r'
r}(t)$, respectively, to Eqs.  (\ref{asymP}) - (\ref{asymA}) at time
separations $t$ and $t+1$. The largest interval at large $t$ showing an
approximate plateau in an effective mass graph we chose as the initial
trial range on which to fit each propatator to the corresponding
asymptotic form of Eqs.  (\ref{asymP}) - (\ref{asymA}).  Similarly,
the largest interval at large $t$ showing an approximate plateau in a
graph of $C^A_{ r' r''}(t) / C^P_{ r' r}(t)$ we chose as the
initial trial range for a fit to Eq. (\ref{asymAoverP}).  
Figures (\ref{fig:570mP}) - (\ref{fig:617mV}) 
show the plateaus at large $t$ in the
effective masses $m^P(t)$, $m^{P'}(t)$, 
the plateau at large $t$ in
the ratio $C^A_{ 0 2}(t) / C^P_{ 0 2}(t)$, and the effective mass
$m^V(t)$ for propagators
with source $r$ of 2 and sink $r'$ of 0. 
Figures (\ref{fig:570mP}) - (\ref{fig:570mV}) show results for the
lattice $16^3 \times 32$ at $\beta$ of 5.70 and the largest
corresponding $k$ value, 0.1675.  Figures (\ref{fig:593mP}) -
(\ref{fig:593mV}) show results for the lattice $24^3 \times 36$ at
$\beta$ of 5.93 and the largest corresponding $k$, 0.1581.  Figures
(\ref{fig:617mP}) - (\ref{fig:617mV}) show results for $30 \times
32^2 \times 40$ at $\beta$ of 6.17 and the largest corresponding $k$, 0.1532.

Fits to data for a range of $t$ were done by minimizing the fit's
$\chi^2$ determined from the full correlation matrix for the data being
fit.  An automatic fitting program repeatedly carried out fits on every
connected interval of four or more points within the initial trial
fitting range. For fits to Eq. (\ref{asymAoverP}), which require only a
single fitting parameter, we looked at intervals of three or more
points.  The final fitting range was chosen by the program to be the
interval with the smallest value of $\chi^2$ per degree of freedom.
Altough a variety of other criteria could be used to determine the final
fitting range, an advantage of the method we adopted is that it can be
implemented automatically thereby reducing the potential for biases.
The reliability of our in our final extrapolated results depends
to some degree on adopting a consistent choice of fitting ranges at
different parameter values.

The horizontal lines in Figures (\ref{fig:570mP}) - (\ref{fig:617mV})
show the fitted values of masses and $Z^A_{0 2} / Z^P_{0 2}$, and the
pair of vertical lines in each figure indicates the interval of $t$
values in the final fitting range chosen. It is perhaps useful to
mention that since the effective mass at each $t$ depends on data both
at $t$ and $t+1$, the effective mass shown at the highest $t$ within
each final fitting range depends on data outside the fitting range.
Thus the fitted lines tend to approximate the average of the effective
masses within the fitting range but with the effective mass at highest
$t$ omitted.  In all but one case, the data shows clear plateaus at
large $t$ extending over more than four time values. These plateaus
appear to be fitted fairly reliably.  For the rho propagator,
$C^V_{02}(t)$, on the lattice $24^3 \times 36$, the plateau is more
ambiguous.  The fit is made over the four $t$ values from 9 to 12, for
which the three corresponding effective masses fall slowly.  The
effective mases at $t$ of 12 to 15 then fall about one standard
deviation below the fit.  At $t$ of 16 and 17, the effective masses then
return to the original fitted value. This behavior is consistent with a
statistical fluctuation although it makes the identication of the
plateau at which to fit more ambiguous.  Comparable ambiguities do not
occur elsewhere in our data.  The interval chosen is approximately a
rescaling of the rho fitting intervals of 5 to 8, at $\beta$ of 5.70 and
13 to 16, at $\beta$ of 6.17.  Also, the rho decay constant obtained
from this fit in Section \ref{sect:decays} is interpolates smoothly
values obtained from less ambiguous fits at other values of quark mass
and lattice spacing.  If this point were simply elimated from our
extrapolations of $f_{\rho}$ in quark mass and in lattice spacing, our
final continuum predictions would be nearly unchanged.

Value of $\chi^2$ for the fits in Figures (\ref{fig:570mP}) -
(\ref{fig:617mV}) are shown in Table
\ref{tab:propchi2}.  Our fits for sinks with size $r'$ of 2, fits at smaller
$k$, and fits on the lattice $24^3 \times 32$ are of comparable quality to
those shown and give comparable $\chi^2$.

Statistical uncertainties of parameters obtained from fits and of any
function of these parameters were determined by the bootstrap method
\cite{Efron}.  From each ensemble of N gauge configurations, 100
bootstrap ensembles were generated. Each bootstrap ensemble consists of
a set of N gauge configurations randomly selected from the underlying N
member ensemble allowing repeats. For each bootstrap ensemble the entire
fit was repeated, including a possibly new choice of the final fitting
interval.  The collection of 100 bootstrap ensembles thus yields a
collection of 100 values of any fitted parameter or any function of any
fitted parameter.  The statistical uncertainty of any parameter is taken
to be half the difference between a value which is higher than all but
15.9\% of the bootstrap values and a value which is lower than all but
15.9\% of the bootstrap values. In the limit of large N the collection
of bootstrap values of a parameter $p$ approaches a gaussian
distribution and the definition we use for statistical uncertainty
approaches the dispersion, $d$, given by $\sqrt{< p^2 > - < p >^2}$.

In the absence of some independent method for determing the predictions
of QCD, it appears inevitable that the choice of $t$ interval on which
to fit data to a large $t$ asymptotic form must be made by some
procedure which depends on the Monte Carlo data itself.  Thus the
statistical uncertainties in the data lead to a corresponding
uncertainty in the choice of fitting interval which, in turn, leads to
some additional uncertainty in the fitted result.  Another advantage of our
procedure for choosing the fitting interval combined with bootstrap
evaluation of statistical uncertainties is that the values we obtain for
statistical uncertainties include the uncertainty arising from the
choice of fitting interval.  A comparison of the error bars found for
our final fits with the error bars found using the same fitting range
held fixed across the bootstrap ensemble shows that typically about 10\%
of the final statistical uncertainty comes from fluctuations over the
bootstrap ensemble of the fitting range itself.

\section{DECAY CONSTANTS}\label{sect:decays}

To construct decay constants using the finite renomalizations of
Eqs. (\ref{defz0}) or (\ref{defz1}), we require values of the critical
hopping constant $k_c$ and of the strong coupling constant 
$\alpha_{\overline{ms}}( 1 / a)$. For $k_c$ we used the values
determined in Ref. \cite{Butler93}. These are listed in Table (\ref{tab:kcrit}).
To determine $\alpha_{\overline{ms}}( 1 / a)$ we first calculated
$\alpha_{\overline{ms}}( \pi / a)$ using the mean-field improved
perturbation theory relation \cite{Lepage,Fermilab}
\begin{eqnarray}
\label{meanfield}
\frac{1}{g^2_{\overline{MS}}( \pi/a)} = \frac{< Tr U / 3>}{g^2_{lat}} + 0.025,
\end{eqnarray}
where $< Tr U>$ is the expectation value of the trace of a plaquette and
$g^2_{lat}$ is $6/\beta$.
We then found $\alpha_{\overline{ms}}( \pi / a)$
given by $ 4 \pi / g^2_{\overline{MS}}( \pi/a) $ and used the two-loop
Callan-Symanzik equation to determine $\alpha_{\overline{ms}}( 1 / a)$.
The corresponding values of $z^{A 1}$ and $z^{V 1}$ are shown
in Table (\ref{tab:zAandzV}).

Values of the decay constant in lattice units $f_{\pi} a$ for the
various lattices, $\beta$ and $k$ shown in Table (\ref{tab:lattices})
are listed in Tables (\ref{tab:f8}) - (\ref{tab:f32}). As a measure of
the lattice spacing in each case, Table (\ref{tab:mrho}) gives the rho
mass in lattice units $m_{\rho}(m_n) a$, extrapolated to the ``normal''
quark mass $m_n$ which produces the physical value of
$m_{\pi}(m_n)/m_{\rho}(m_n)$ \cite{Butler93}. We list also values of
$m_n$ itself. 
These
parameters are not given for the lattice $8^3 \times 32$ since in this
case we were not able to calculate propagators at small enough quark
mass to perform the required extrapolation reliably.  The finite
renormalizations for the decay constants in Tables (\ref{tab:f8}) -
(\ref{tab:f32}) all include both the leading term $z^{A 0}$ of Eqs.
(\ref{defz0}) and the first order mean-field improved perturbative
correction $z^{A 1}$ of Eqs. (\ref{defz1}).  The second column in each
table gives values of $f_{\pi} a$ found from $Z^A_{ r' r}$
determined from a direct fit of $C^A_{ r' r}( t)$ to Eq.
(\ref{asymA}). The third column gives $f_{\pi} a$ found from $Z^A_{
r' r}$ determined by fitting the ratio $C^A_{ r' r}( t) / C^P_{
r' r}( t)$ to Eq. (\ref{asymAoverP}) and then using the value of
$Z^P_{ r' r}$ found from a fit of $C^P_{ r' r}( t)$ to Eq.
(\ref{asymP}). The two sets of data in all cases are statistically
consistent and in all cases, except for the lattice $8^3 \times 32$ for
$k$ below 0.1500, $f_{\pi}$ determined from ratio fits has a smaller statistical error
$f_{\pi}$ determined from direct fits.
The ratio method tends to give less statistical
noise, in effect, because it uses $C^P_{ r' r}( t)$, which is
relatively less noisey, to determine $m^P$ and then extracts only
$Z^A_{ r' r}$ from the more noisey propagator $C^A_{ r' r}( t)$.
The direct method extracts both $Z^A_{ r' r}$ and $m^P$ from
$C^A_{ r' r}( t)$ yielding an $m^P$ with greater noise, which is
then multiplied by a possibly large $t$ and exponentiated, feeding
additional noise back into the value of $Z^A_{ r' r}$. In the
remainder of this paper we use only values of $f_{\pi}$ determined from
ratio fits.

Values of the decay constant in lattice units $F_{\rho} a$ for the
lattices shown in Table (\ref{tab:lattices}) are listed in the fourth
column of Tables (\ref{tab:f8}) - (\ref{tab:f32}).  The finite
renormalizations for $F_{\rho}$ shown in these tables all include both
the leading term $z^{V 0}$ of Eqs. (\ref{defz0}) and the first order
mean-field improved perturbative correction $z^{V 1}$ of Eqs.
(\ref{defz1}).  The values of $Z^V_{ r' r}$ used to determine
$F_{\rho}$ were all extracted from direct fits of $C^V_{ r' r}( t)$
to Eq. (\ref{asymV}).

\section{VOLUME DEPENDENCE}\label{sect:vol}

Percentage changes in decay constants going from $8^3 \times 32$ and
$16^3 \times 32$ to $24^3 \times 32$, at $\beta$ of 5.70, are given in
Table~\ref{tab:voldep}.  These changes are the same for all choices of
finite renormalization.  All of the differences appear to be of marginal
statistical significance and may therefore best be viewed as upper
bounds on the volume dependence of our results.  A variety of different
arguments suggest that, for the range of $k$, $\beta$, and lattice
volume we have examined, the errors in valence approximation decay
constants due to calculation in a finite volume $L^3$ are bounded by an
expression of the form $C e^{- L/R}$.  A simple non-relativistic
potential model gives this expression with the radius of a hadron's wave
function for $R$. A more elaborate field theory argument gives for $R$
the Compton wave length of a pair of pions, which is the lightest state
that can be exchanged between a pair of identical pseudoscalar or vector
mesons.  At $\beta$ of 5.70, $R$ is thus very likely to be between 3 and
5 lattice units. Since the changes in decay constants shown in
Table~\ref{tab:voldep} between $16^3$ and $24^3$ are all less than 5\%
for k $\ge 0.1650$, it follows that the differences between these values
in $24^3$ and in infinite volume should be less than 1.3\%.  For
extrapolations to physical quark masses, we use only decay constants
with $k \ge 0.1650$.

\section{QUARK MASS EXTRAPOLATION}\label{sect:mextrap}

At the largest $k$ on each lattice, the ratio $m_{\pi} / m_{\rho}$ is
significantly larger than its experimentally observed value of 0.179.
Thus to produce decay constants for hadrons containing only light
quarks, our data has to be extrapolated to larger $k$ or, equivalently,
to smaller quark mass.  We did not calculate directly at larger $k$ both
because the algorithms we used to find quark propagators became too slow
and because the statistical errors we found in trial calculations became
too large.

Define the
quark mass in lattice units $m_q a$ to be
\begin{eqnarray}
\label{defmq}
m_q a = \frac{1}{2 k} - \frac{1}{2 k_c},
\end{eqnarray}
where $k_c$ is the critical hopping constant at which $m_{\pi}$ becomes zero.
We found $f_{\pi} a$ and $F_{\rho} a$, for all three possible choices of
finite renormalization in Eqs. (\ref{defnaive}) - (\ref{defz1}) to be
nearly linear functions of $m_q a$ over the entire range of $k$
considered on each lattice.

\begin{figure}
\epsfxsize=\textwidth \epsfbox{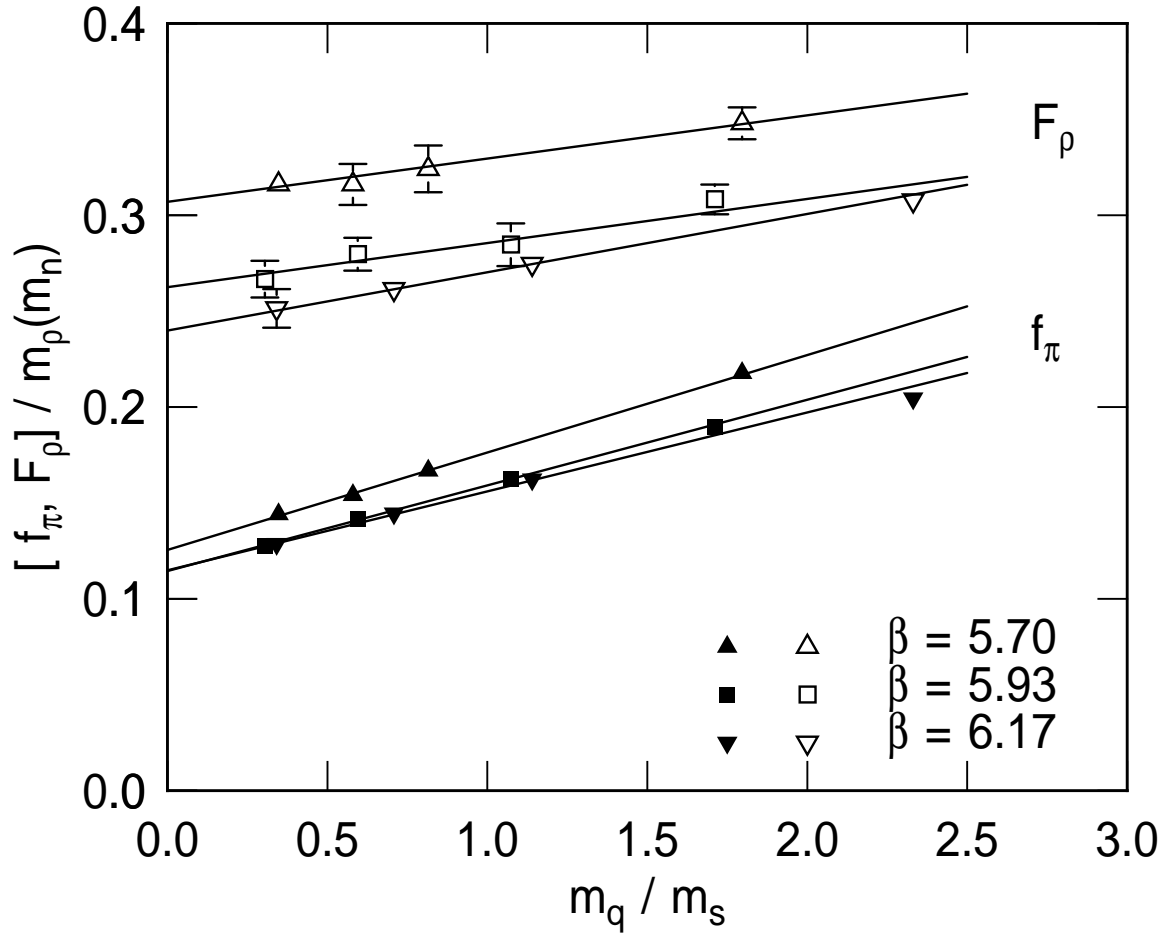}
\caption{ The decay constants $F_{\rho}$ and $f_{\pi}$, with perturbative
renormalization, in units of the central value of the physical rho mass $m_{\rho}(m_n)$, as
functions of the quark mass $m_q$, in units of the central value of the strange quark mass
$m_s$.
\label{fig:mextrap}}
\end{figure}

Figure~\ref{fig:mextrap} shows $f_{\pi}$ and $F_{\rho}$, with finite
renormalizations including the first order mean-field improved
perturbative correction, as functions of $m_q$.  Data is shown from the
three lattices of Table~\ref{tab:lattices} which we use to evaluate
continuum limits, $16^3 \times 32$, $24^3 \times 36$ and $30 \times 32^2
\times 40$.  For convenience, decay constants at each $\beta$ are shown
in units of the central value of the rho mass $m_{\rho}(m_n)$, at the same
$\beta$, extrapolated to the the ``normal'' quark mass $m_n$.  Quark
masses $m_q$ for each $\beta$ are shown in units of the central value of
the strange quark mass $m_s$ at the same $\beta$. The data 
Figure~\ref{fig:mextrap} has been scaled by the central values of $m_{\rho}(m_n)$
and $m_s$ taken as arbitrary external parameters, and the error bars
shown do not include the effect of statistical fluctuations in  
$m_{\rho}(m_n)$ or $m_s$.
Table
(\ref{tab:mrho}) gives values of $m_s$ found in Ref. \cite{Butler93} by
requiring $m_{\pi}[ (m_n + m_s)/2] / m_{\rho}(m_n)$ to be equal to the
physical value of $m_K/m_{\rho}$.  The lines in Figure~\ref{fig:mextrap}
are fits of
decay constants measured in lattice units, $f_{\pi} a$ and $F_{\rho} a$,
to linear functions of the quark mass in lattice units, $m_q a$, at the
three smallest quark masses in the data set at each $\beta$.  These fits
were obtained by minimizing $\chi^2$ obtained form the full correlation
matrix amoung the data points. The correlation matrix was calculated by
taking averages of data values and products of data values over 
bootstrap ensembles generated as described in Section~(\ref{sect:props}).  The
$\chi^2$ for these fits, and corresponding fits on the lattice $24^3
\times 32$ at $\beta$ of 5.70, are given in Table \ref{tab:mqchi2}.  The
fits in Figure~\ref{fig:mextrap} appear to be quite good and provide, we
believe, a reliable method for extrapolating decay constants down to
light quark masses.  
With naive finite renormalization, Eq.
(\ref{defnaive}), or zero order mean-field finite renormalization, Eq.
(\ref{defz0}), $f_{\pi} a$ and $F_{\rho} a$ fit straight lines in $m_q a$
about as well as the first order perturbatively renormalized data of
Figure~\ref{fig:mextrap}. 

The linear fits of Figure~\ref{fig:mextrap} permit the determination of
$f_K$ and $F_{\phi}$ in addition to $f_{\pi}$ and $F_{\rho}$.  For a
pion composed of a quark and antiquark with mass $m_q \neq
m_{\overline{q}}$, Figure~\ref{fig:mextrap} suggests
\begin{eqnarray}
\label{fpiofmq}
f_{\pi} = \alpha_q m_q + \alpha_{\overline{q}} m_{\overline{q}} + \beta.  
\end{eqnarray}
Charge conjugation invariance then gives $\alpha_q =
\alpha_{\overline{q}}$.  It follows that the kaon, which is a pion with,
say, $m_q = m_s$ and $m_{\overline{q}} = m_n$, will have the same decay
constant as a pion composed of a single type of quark and antiquark with
$m_q = m_{\overline{q}} = (m_s + m_n)/2$.  On the other hand, the linear
fits of Figure~\ref{fig:mextrap} permit $F_{\rho}$ to be extrapolated to
the point $m_q = m_{\overline{q}} = m_s$ which, in the valence
approximation, gives $F_{\phi}$. Tables \ref{tab:fPofa} and 
\ref{tab:fVofa} give the value
of $f_{\pi} a$, $f_K
a$, $F_{\rho} a$ and $F_{\phi} a$ 
obtained from the fits in Figure~\ref{fig:mextrap}.
The statistical uncertainties in
these quantities were obtained by a further application of the bootstrap
method of Section~(\ref{sect:props}). Bootstrap ensembles of the
underlying gauge configurations were generated, and on each bootstrap
ensemble the extrapolated decay constants were recalculated. 
The uncertainty in each decay constant was obtained from the resulting
distribution. The correlation matrices used to fit bootstrap data to
linear functions of $m_q a$ were taken to be the same as the correlation
matrices for the full ensemble. To
recalulate correlation matrices separately on each bootstrap ensemble by a further
bootstrap would have been too time consuming.

\section{CONTINUUM LIMIT}\label{sect:contlim}

\begin{figure}
\epsfxsize=\textwidth \epsfbox{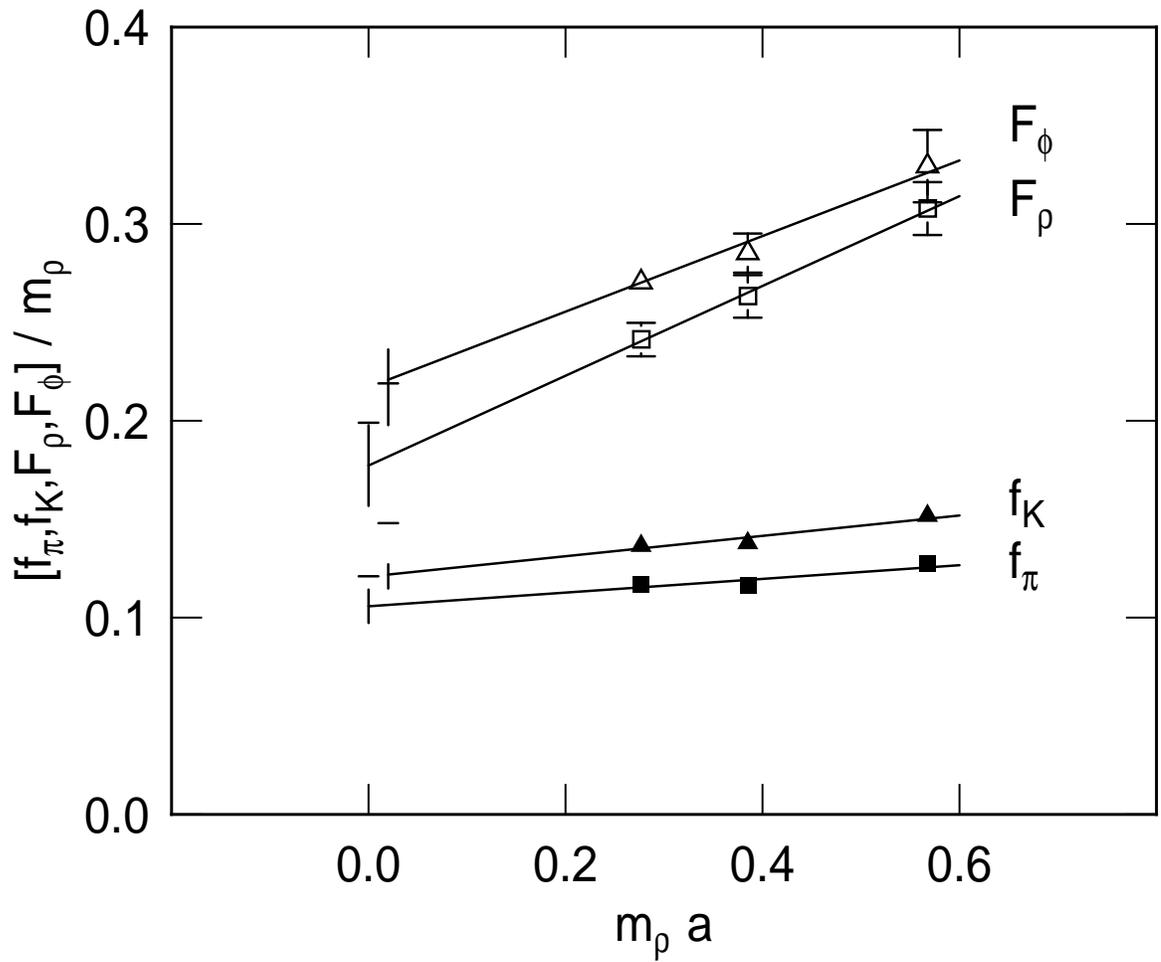}
\caption{ Perturbatively renormalized 
decay constants as functions of the lattice
spacing $a$, in units of $1/m_{\rho}$. The error bars near zero lattice
spacing are uncertainties in the extrapolated ratios, and the horizontal
lines represent experimentally observed values.
\label{fig:aextrap}}
\end{figure}

The ratios $f_{\pi} / m_{\rho}$, $f_K / m_{\rho}$, $F_{\rho} / m_{\rho}$
and $F_{\phi} / m_{\rho}$ for physical quark masses we then extrapolated
to zero lattice spacing.  For Wilson fermions the leading asymptotic
lattice spacing dependence in these decay ratios is expected to be
linear in $a$.  On the other hand, as shown in Ref.~\cite{Butler93},
$m_{\rho} a$ follows the two-loop Callan-Symanzik scaling prediction in
$\alpha_{\overline{ms}}$ quite well for the range of $\beta$ considered
here. Thus assuming the asymptotic form of the lattice spacing
dependence of decay ratios appears to be reasonable.
Figure~\ref{fig:aextrap} shows decay constants with first order perturbative
renormalization along with fits to linear functions of $m_{\rho} a$.
The quantity $m_{\rho} a$ may be viewed as the lattice spacing $a$
measured in units of the physical rho Compton wavelength, $1/m_{\rho}$.
The vertical bars at $m_{\rho} a$ of 0, offset slightly for visibility,
are the extrapolated predictions' uncertainties.
The horizontal lines at $m_{\rho} a$ of 0 lying within or slightly above
the range of each prediction are the corresponding experimental values.
The data points in Figure~\ref{fig:aextrap} are from the lattices $16^3
\times 32$, $24^3 \times 36$ and $30 \times 32^2
\times 40$.  The values of $\beta$ for these lattices were chosen so
that the physical volume in each case is nearly the same.  For lattice
period L, the quantity $m_{\rho} L$ is respectively, 9.08 $\pm$ 0.13,
9.24 $\pm$ 0.19 and, averaged over three directions, 8.67 $\pm$ 0.12
\cite{Butler93}.  

The fits shown in Figure~\ref{fig:aextrap} were found by minimizing
$\chi^2$ obtained from the full correlation matrix among the fitted
data.  Since both the x and y coordinates of each of the three fitted
points on each line have statistical uncertainties, we evaluated
$\chi^2$ among all six pieces of data and chose as fitting parameters
the slope and intercept of the line along with the x coordinate of each
point.  The required correlation matrices were found by the bootstrap
method as were the statistical uncertainties of the extrapolated
predictions.  The correlation matrices used in fits for each bootstrap
ensemble were again taken from the full ensemble and not recalculated on
each bootstrap ensemble independently.  The $\chi^2$ per degree of
freedom for the fits in Figure~\ref{fig:aextrap}, are given in
Table~\ref{tab:slopechi}. 

For the lattice spacing dependence of decay ratios found using zeroth
order mean-field finite renormalization and using naive finite
renormalization we also made fits to linear functions of $m_{\rho} a$.
The slopes with respect to $m_{\rho a}$ of the ratios $f_{\pi} /
m_{\rho}$, $f_K / m_{\rho}$, $F_{\rho} / m_{\rho}$ and $F_{\phi} /
m_{\rho}$ along with $\chi^2$ for each fit are also given in
Table~\ref{tab:slopechi}.  It is clear from these results, as
menitioned in Section \ref{sect:defs}, that naive renormalization leads
to decay ratios with significantly stronger lattice spacing dependence
than found for either zeroth or first order improved perturbative
renormalization. There also appears to be some tendency for first order
perturbative renormalization to lead to weaker lattice spacing
dependence than zeroth order. It follows that the extrapolations we have
done to zero lattice spacing are likely to be most reliable for first
order perturbative renormalzation and least reliable for naive
renormalization.

\section{INFINITE VOLUME LIMIT}

The continuum ratios we found in finite volume were then corrected to
infinite volume by an adaptation of the method used in Ref.
\cite{Butler93} to correct finite volume continuum mass ratios to
infinite volume. From $f_{\pi} / m_{\rho}$, with first order
perturbative renormalization, 
as a function of lattice spacing $a$ and lattice period $L$, both
measured in physical units, define the finite volume correction term
$\Delta( a, L)$ to be
\begin{eqnarray}
\label{defdelta}
\Delta( a, L) = \frac{f_{\pi}}{m_{\rho}}( a, \infty) - 
\frac{f_{\pi}}{m_{\rho}}( a, L).
\end{eqnarray}
The quantity which we would like to determine
is $\Delta( 0, 9/m_{\rho})$.
Now the ratio $f_{\pi} / m_{\rho}$,
for $L$ of $9/m_{\rho}$,
undergoes a relative change of a bit less than 20\% as $a$ goes from its value 
$a_{5.7}$ at
$\beta$ of 5.70 to 0.
Thus we would expect an error of about 20\% of $\Delta( 0, 9/m_{\rho})$
for the approximation
\begin{eqnarray}
\label{approxina}
\Delta( 0, \frac{9}{m_{\rho}}) \approx \Delta( a_{5.7}, \frac{9}{m_{\rho}}).
\end{eqnarray}
Moreover, from our earlier discussion 
of the exponential approach of decay ratios to their infinite volume limits,
it follows that
with an additional
error of about 20\% of $\Delta( 0, 9/m_{\rho})$ we have
\begin{eqnarray}
\label{approxinL}
\Delta( a_{5.7}, \frac{9}{m_{\rho}}) \approx 
\frac{f_{\pi}}{m_{\rho}}( a_{5.7}, \frac{13.5}{m_{\rho}}) - 
\frac{f_{\pi}}{m_{\rho}}( a_{5.7}, \frac{9}{m_{\rho}}), 
\end{eqnarray}
where $L$ of $13.5 / m_{\rho}$ corresponds to the lattice
$24^3 \times 32$ at $\beta$ of 5.70.
Finally, a direct evaluation of the
right side of Eq.~(\ref{approxinL}) shows it is 
quite small, with statistical errors of
about 6\% of $f_{\pi}/m_{\rho}$. 
Combining Eqs. (\ref{defdelta}) - (\ref{approxinL}), we obtain
\begin{eqnarray}
\label{approxfinal}
\frac{f_{\pi}}{m_{\rho}}( 0, \infty) \approx
\frac{f_{\pi}}{m_{\rho}}( 0, \frac{9}{m_{\rho}}) + 
\frac{f_{\pi}}{m_{\rho}}( a_{5.7}, \frac{13.5}{m_{\rho}}) - 
\frac{f_{\pi}}{m_{\rho}}( a_{5.7}, \frac{9}{m_{\rho}}). 
\end{eqnarray}
The error in this approximation should be less than about 40\% of 6\% of
$f_{\pi}/m_{\rho}$, which is 2.4\% of $f_{\pi}/m_{\rho}$.  Table
\ref{tab:syserrs} lists estimates of the systematic uncertainties in
equations corresponding to Eq. (\ref{approxfinal}) for other decay
constants and other choices of finite renormalization. First order
perturbative renormalization consistently gives the smallest systematic
error in volume correction largely because the lattice spacing
dependence of these decay ratios is smallest.

The ratios $f_{\pi} / m_{\rho}$, $f_K / m_{\rho}$, $F_{\rho} / m_{\rho}$
and $F_{\phi} / m_{\rho}$, for all three different choices of
finite renormalization, extrapolated to zero lattice spacing with
$m_{\rho} L$ fixed at 9, and then corrected to infinite volume are shown
in Table~\ref{tab:res}.  For first order perturbative renormalization
we also give, in Table~\ref{tab:resoverfk}, finite and infinite volume
values of the ratios $f_{\pi} / f_K$, $F_{\rho} / f_K$
and $F_{\phi} / f_K$.
The errors shown for infinite volume ratios
are statistical only and do not include the estimates we have just given
for the systematic error in our procedure for making infinite volume
corrections.  
For first order perturbative finite
renormalization, a comparision of Tables \ref{tab:syserrs} and
\ref{tab:res} shows, however, that the systematic errors arising from our
method of obtaining infinite volume results are
much smaller than the statistical errors.  For the range of
$\beta$ used in our extrapolation to zero lattice spacing, first order
mean-field theory improved perturbation expansions have been
shown~\cite{Lepage} to work quite well for a wide variety of different
quantities. In addition, as we mentioned earlier, the extrapolation to zero lattice
spacing should be most reliable for this renormalization scheme.
Thus we believe that the numbers in Table~\ref{tab:res}
obtained using first order mean-field perturbative renormalization are
significantly more reliable than those found using the other two
renormalization methods.  Zeroth order perturbative renormalization
has been included,
however, to provided some measure of the degree to which our results may
be sensitive to the choice of renormalization. Half of the difference
between first order and zeroth order mean-field perturbative
renormalization appears to us to be a conservative estimate of the
systematic uncertainty in the first order results arising from the
missing second and higher order perturbative renormalization contributions.
In all cases this uncertainty is significantly less than the statistical
errors. 
Predictions obtained with naive renormalization have been included in 
Table \ref{tab:res} largely as a curiosity. It is 
interesting to notice, however, that the difference between
the final, infinite volume results found with naive 
renormalization and those found with first order perturbative
renormalization is still less than 1.5 times the naive renormalization 
statistical errors.

The predicted infinite volume ratios in Table~\ref{tab:res} are all
statistically consistent with the corresponding finite volume ratios.
The main consequence of the correction to infinite volume is an increase
in the size of the statistical uncertainty in each prediction.

The experimental numbers shown in Table~\ref{tab:res} for $f_{\pi}$ and
$f_K$ are from charged particle decays and for $f_{\rho}$ from neutral
decays. In all cases the uncertainties in the experimental values are
0.001 or less. As mentioned in the introduction, an experimental value
for the neutral pion decay \cite{Cello} gives $f_{\pi} / m_{\rho}$ of
$0.110 \pm 0.005$, which is quite close to our prediction. The
systematic uncertainties in this experimental number, however, are
larger than those for the charged pion decay. As a result the
significance of the improved agreement of our prediction with the
observed neutral pion decay constant is unclear to us.

We would like to thank Paul Mackenzie for discussions, and Mike Cassera,
Molly Elliott, Dave George, Chi Chai Huang and Ed Nowicki for their work
on GF11. We are particularly grateful to Chris Sachrajda for calling our
attention to an error in an earlier version of this paper.

\newpage

\begin{table}
\begin{center}
\begin{tabular}{r@{}lcccc}     \hline
 \multicolumn{2}{c}{lattice}          & $\beta$ & k     & skip & count\\ 
 \hline
 $8^3$ & $\:\times \: 32$ & 5.70 & 0.1400 - 0.1650 & 1000  & 2439\\ 
 $16^3$ & $\:\times \: 32$ & 5.70 & 0.1600 - 0.1675 & 2000 & 219\\ 
 $24^3$ & $\:\times \: 32$ & 5.70 & 0.1600 - 0.1675 & 4000 & 58\\ 
 $24^3$ & $\:\times \: 36$ & 5.93 & 0.1543 - 0.1581 & 4000 & 210\\
 $30 \times 32^2$ & $\:\times \: 40$ & 6.17 & 0.1500 - 0.1532 & 6000 & 219\\
 \hline
\end{tabular}
\caption{Configurations analyzed.}
\label{tab:lattices}
\end{center}
\end{table}

\begin{table}
\begin{center}
\begin{tabular}{r@{}lccccc}     \hline
 \multicolumn{2}{c}{lattice}          & $\beta$ & $C^P_{02}$ & $C^A_{02}$ & 
                                         $C^A_{02} / C^P_{02}$ & $C^V_{02}$ \\
 \hline
 $16^3$ & $\:\times \: 32$ & 5.70 & 0.09 & 0.22 & 0.23 & 0.11 \\
 $24^3$ & $\:\times \: 36$ & 5.93 & 0.08 & 0.07 & 0.47 & 0.68 \\
 $30 \times 32^2$ & $\:\times \: 40$ & 6.17 & 0.16 & 0.10 & 0.01 & 0.25 \\
 \hline
\end{tabular}
\caption{Values of $\chi^2$ per degree of freedom for fits to propagators.}
\label{tab:propchi2}
\end{center}
\end{table}

\begin{table}
\begin{center}
\begin{tabular}{r@{}lcc}     \hline
 \multicolumn{2}{c}{lattice}          & $\beta$ & $k_c$ \\ 
 \hline
 $8^3$ & $\:\times \: 32$ & 5.70 & $0.169012 \pm 0.000102$ \\ 
 $16^3$ & $\:\times \: 32$ & 5.70 & $0.169405 \pm 0.000052$ \\ 
 $24^3$ & $\:\times \: 32$ & 5.70 & $0.169304 \pm 0.000035$ \\ 
 $24^3$ & $\:\times \: 36$ & 5.93 & $0.158948 \pm 0.000026$\\
 $30 \times 32^2$ & $\:\times \: 40$ & 6.17 & $0.153763 \pm 0.000018$\\
 \hline
\end{tabular}
\caption{Critical hopping constant values.}
\label{tab:kcrit}
\end{center}
\end{table}

\begin{table}
\begin{center}
\begin{tabular}{ccc}     \hline
  $\beta$ & $z^{A 1}$ & $z^{V 1}$\\ 
 \hline
   5.70 &  0.93269 & 0.82194  \\ 
   5.93 &  0.94208 & 0.84679  \\
   6.17 &  0.94828 & 0.86320  \\
 \hline
\end{tabular}
\caption{First order perturbative correction factors to finite renormalizations.}
\label{tab:zAandzV}
\end{center}
\end{table}

\begin{table}
\begin{center}
\begin{tabular}{cccc}     \hline
 k   & $f_{\pi} a$ direct & $f_{\pi} a$ ratio & $F_{\rho} a$ \\ \hline
 0.1400 &  $0.2346 \pm 0.0011$ & $0.2347 \pm 0.0011$ & $0.2861 \pm 0.0015$\\
 0.1450 &  $0.2080 \pm 0.0014$ & $0.2083 \pm 0.0010$ & $0.2684 \pm 0.0014$\\
 0.1500 &  $0.1811 \pm 0.0016$ & $0.1820 \pm 0.0010$ & $0.2505 \pm 0.0017$\\
 0.1550 &  $0.1532 \pm 0.0018$ & $0.1538 \pm 0.0012$ & $0.2316 \pm 0.0026$\\ 
 0.1600 &  $0.1238 \pm 0.0022$ & $0.1243 \pm 0.0012$ & $0.2074 \pm 0.0022$\\       
 0.1650 &  $0.0895 \pm 0.0074$ & $0.0925 \pm 0.0021$ & $0.1843 \pm 0.0043$\\ 
\hline      
\end{tabular}
\caption{Decay constants in lattice units measured on a lattice $8^3 \times 32$ 
at $\beta = 5.70$. Finite renormalizations include the first order
perturbative correction.}
\label{tab:f8}
\end{center}
\end{table}

\clearpage

\begin{table}
\begin{center}
\begin{tabular}{cccc}     \hline
 k   & $f_{\pi} a$ direct & $f_{\pi} a$ ratio & $F_{\rho} a$ \\ \hline
 0.1600  &  $0.1212 \pm 0.0040$ & $0.1236 \pm 0.0016$ & $0.1975 \pm 0.0047$\\
 0.1650  &  $0.0915 \pm 0.0046$ & $0.0947 \pm 0.0019$ & $0.1840 \pm 0.0069$\\
 0.16625 &  $0.0799 \pm 0.0061$ & $0.0875 \pm 0.0022$ & $0.1794 \pm 0.0061$\\
 0.1675  &  $0.0720 \pm 0.0098$ & $0.0818 \pm 0.0025$ & $0.1794 \pm 0.0034$\\
\hline      
\end{tabular}
\caption{Decay constants in lattice units measured
on a lattice $16^3 \times 32$ at $\beta = 5.70$.
Finite renormalizations include the first order 
perturbative correction.}
\label{tab:f16}
\end{center}
\end{table}

\begin{table}
\begin{center}
\begin{tabular}{cccc}     \hline
 k   & $f_{\pi} a$ direct & $f_{\pi} a$ ratio & $F_{\rho} a$ \\ \hline
 0.1600  &  $0.1251 \pm 0.0026$ & $0.1249 \pm 0.0010$ & $0.2116 \pm 0.0035$\\
 0.1650  &  $0.0951 \pm 0.0031$ & $0.0940 \pm 0.0010$ & $0.1930 \pm 0.0055$\\ 
 0.1663  &  $0.0792 \pm 0.0083$ & $0.0860 \pm 0.0015$ & $0.1865 \pm 0.0052$\\       
 0.1675  &  $0.0769 \pm 0.0059$ & $0.0787 \pm 0.0018$ & $0.1752 \pm 0.0043$\\ 
\hline      
\end{tabular}
\caption{Decay constants in lattice units measured 
on a lattice $24^3 \times 32$ at $\beta = 5.70$.
Finite renormalizations include the first order 
perturbative correction.}
\label{tab:f24b57}
\end{center}
\end{table}

\begin{table}
\begin{center}
\begin{tabular}{cccc}     \hline
 k   & $f_{\pi} a$ direct & $f_{\pi} a$ ratio & $F_{\rho} a$ \\ \hline
 0.1543  &  $0.0763 \pm 0.0025$ & $0.0729 \pm 0.0009$ & $0.1187 \pm 0.0030$\\
 0.1560  &  $0.0668 \pm 0.0037$ & $0.0625 \pm 0.0008$ & $0.1096 \pm 0.0043$\\ 
 0.1573  &  $0.0562 \pm 0.0021$ & $0.0544 \pm 0.0009$ & $0.1077 \pm 0.0033$\\       
 0.1581  &  $0.0517 \pm 0.0029$ & $0.0490 \pm 0.0012$ & $0.1027 \pm 0.0037$\\ 
\hline      
\end{tabular}
\caption{
Decay constants in lattice units measured 
on a lattice $24^3 \times 36$ at $\beta = 5.93$.
Finite renormalizations include the first order
perturbative correction.}
\label{tab:f24b59}
\end{center}
\end{table}

\begin{table}
\begin{center}
\begin{tabular}{cccc}     \hline
 k   & $f_{\pi} a$ direct & $f_{\pi} a$ ratio & $F_{\rho} a$ \\ \hline
 0.1500  &  $0.0557 \pm 0.0009$ & $0.0566 \pm 0.0005$ & $0.0852 \pm 0.0008$\\
 0.1519  &  $0.0444 \pm 0.0012$ & $0.0449 \pm 0.0004$ & $0.0760 \pm 0.0010$\\ 
 0.1526  &  $0.0403 \pm 0.0017$ & $0.0400 \pm 0.0005$ & $0.0724 \pm 0.0013$\\       
 0.1532  &  $0.0361 \pm 0.0023$ & $0.0356 \pm 0.0007$ & $0.0696 \pm 0.0028$\\ 
\hline      
\end{tabular}
\caption{
Decay constants in lattice units measured 
on a lattice $30 \times 32^2 \times 40$ at $\beta = 6.17$.
Finite renormalizations include the first order
perturbative correction.}
\label{tab:f32}
\end{center}
\end{table}

\begin{table}
\begin{center}
\begin{tabular}{r@{}lcccc}     \hline
 \multicolumn{2}{c}{lattice}          & $\beta$ & $m_{\rho} a$ & $m_n a$
& $m_s a$\\ 
 \hline
 $16^3$ & $\:\times \: 32$ & 5.70 & $0.5676 \pm 0.0079$ &
       $0.00390 \pm 0.00012$ & $0.09662 \pm 0.00291 $ \\ 
 $24^3$ & $\:\times \: 32$ & 5.70 & $0.5409 \pm 0.0089$ &
       $0.00348 \pm 0.00012$ & $0.08620 \pm 0.00303$ \\ 
 $24^3$ & $\:\times \: 36$ & 5.93 & $0.3851 \pm 0.0079$ &
       $0.00223 \pm 0.00009$ & $0.05536 \pm 0.00226$ \\
 $30 \times 32^2$ & $\:\times \: 40$ & 6.17 & $0.2768 \pm 0.0039$ &
       $0.00141 \pm 0.00004$ & $0.03503 \pm 0.00096$ \\
 \hline
\end{tabular}
\caption{The normal and strange quark masses, $m_n$ and
$m_s$, and the rho mass extrapolated to quark mass $m_n$.}
\label{tab:mrho}
\end{center}
\end{table}

\begin{table}
\begin{center}
\begin{tabular}{crrr}     \hline
 decay   & k & $f(24^3) - f(8^3)$ & $f(24^3) - f(16^3)$ \\ \hline
$f_{\pi}$ & 0.1600  & $0.4 \pm 1.4 \% $ & $ 1.0 \pm 1.5 \%$ \\
          & 0.1650  & $1.7 \pm 2.6 \% $ & $-0.7 \pm 2.3 \%$ \\
          & 0.1663  &                   & $-1.7 \pm 3.3 \%$ \\
          & 0.1675  &                   & $-4.1 \pm 4.1 \%$ \\ \hline
$F_{\rho}$ & 0.1600 & $2.0 \pm 1.7 \% $ & $ 6.7 \pm 2.9 \%$ \\       
          & 0.1650  & $4.5 \pm 3.9 \% $ & $ 4.7 \pm 5.1 \%$ \\       
          & 0.1663  &                   & $ 3.8 \pm 4.6 \%$ \\       
          & 0.1675  &                   & $-2.4 \pm 2.8 \%$ \\ \hline
\end{tabular}
\caption{
Changes in decay constants from 
$8^3 \times 32$ to $24^3 \times 32$ and from
$16^3 \times 32$ to $24^3 \times 32$ 
at $\beta = 5.70$}
\label{tab:voldep}
\end{center}
\end{table}

\begin{table}
\begin{center}
\begin{tabular}{r@{}lccc}     \hline
 \multicolumn{2}{c}{lattice}          & $\beta$ & 
                           $\chi^2$ for $f_{\pi}$ & $\chi^2$ for $F_{\rho}$ \\ 
 \hline
 $16^3$ & $\:\times \: 32$ & 5.70 &  0.256 & 0.149 \\ 
 $24^3$ & $\:\times \: 32$ & 5.70 &  0.003 & 0.180 \\ 
 $24^3$ & $\:\times \: 36$ & 5.93 &  0.394 & 0.358 \\
 $30 \times 32^2$ & $\:\times \: 40$ & 6.17 &0.130 & 0.014 \\
 \hline
\end{tabular}
\caption{Values of $\chi^2$ for fits of decay constants to linear
functions of $m_q$ at the three smallest values of $m_q$.}
\label{tab:mqchi2}
\end{center}
\end{table}

\clearpage

\begin{table}
\begin{center}
\begin{tabular}{r@{}lccc}     \hline
 \multicolumn{2}{c}{lattice}          & $\beta$ & 
        $f_{\pi} a$ & $f_K a$ \\ 
 \hline
 $16^3$ & $\:\times \: 32$ & 5.70 & $0.0724 \pm  0.0030$ 
                                  & $0.0861 \pm  0.0017$ \\
 $24^3$ & $\:\times \: 32$ & 5.70 & $0.0691 \pm  0.0025$ 
                                  & $0.0831 \pm  0.0016$ \\
 $24^3$ & $\:\times \: 36$ & 5.93 & $0.0449 \pm  0.0012$ 
                                  & $0.0532 \pm  0.0008$ \\
 $30 \times 32^2$ & $\:\times \: 40$ & 6.17 &$0.0323 \pm  0.0008$ 
                                  & $0.0378 \pm  0.0005$ \\
 \hline
\end{tabular}
\caption{Decay constants in lattice units extrapolated to
physical quark mass.}
\label{tab:fPofa}
\end{center}
\end{table}

\begin{table}
\begin{center}
\begin{tabular}{r@{}lccc}     \hline
 \multicolumn{2}{c}{lattice}          & $\beta$ & 
        $F_{\rho} a$ & $F_{\phi} a$ \\ 
 \hline
 $16^3$ & $\:\times \: 32$ & 5.70           & $ 0.1747 \pm  0.0076  $
                                            & $ 0.1870 \pm  0.0107  $ \\
 $24^3$ & $\:\times \: 32$ & 5.70           & $ 0.1642 \pm  0.0070  $ 
                                            & $ 0.1975 \pm  0.0063  $ \\
 $24^3$ & $\:\times \: 36$ & 5.93           & $ 0.1014 \pm  0.0047  $
                                            & $ 0.1098 \pm  0.0037  $ \\
 $30 \times 32^2$ & $\:\times \: 40$ & 6.17 & $ 0.0668 \pm  0.0026  $
                                            & $ 0.0748 \pm  0.0012  $ \\
 \hline
\end{tabular}
\caption{Decay constants in lattice units extrapolated to
physical quark mass.}
\label{tab:fVofa}
\end{center}
\end{table}

\begin{table}
\begin{center}
\begin{tabular}{clll}     \hline
 decay  & renorm. & slope     & $\chi^2$ \\ \hline
$f_{\pi} / m_{\rho}$ & first order perturb.  & $0.035 \pm 0.022$ &
                                               0.62               \\
                     & zeroth order perturb. & $0.044 \pm 0.023$ &         
                                               0.61               \\
                     & naive                 & $0.112 \pm 0.030$ & 
                                               0.74               \\ \hline
$f_K / m_{\rho}$     & first order perturb.  & $0.052 \pm 0.016$ & 
                                               1.04               \\
                     & zeroth order perturb. & $0.063 \pm 0.017$ &                 
                                               1.02               \\       
                     & naive                 & $0.112 \pm 0.022$ &                 
                                               0.99               \\ \hline
$F_{\rho}/ m_{\rho}$ & first order perturb.  & $0.228 \pm 0.047$ &
                                               0.04               \\ 
                     & zeroth order perturb. & $0.325 \pm 0.058$ &                 
                                               0.06               \\       
                     & naive                 & $0.546 \pm 0.075$ &                 
                                               0.19               \\ \hline
$F_{\phi}/ m_{\rho}$ & first order perturb.  & $0.192 \pm 0.061$ &  
                                               0.20               \\ 
                     & zeroth order perturb. & $0.286 \pm 0.073$ &                 
                                               0.22               \\       
                     & naive                 & $0.362 \pm 0.091$ &                 
                                               0.20               \\ \hline
\end{tabular}
\caption{
Slope and $\chi^2$ of fits of decay ratios' lattice spacing dependence
to linear functions of $m_{\rho} a$.}
\label{tab:slopechi}
\end{center}
\end{table}

\begin{table}
\begin{center}
\begin{tabular}{clr}     \hline
 decay  & renorm. & error \\ \hline
$f_{\pi} / m_{\rho}$ & first order perturb.  & 2\% \\
                     & zeroth order perturb.  & 3\% \\
                     & naive               & 4\% \\
\hline
$f_K / m_{\rho}$     & first order perturb.  & 2\% \\
                     & zeroth order perturb.  & 2\% \\
                     & naive               & 2\% \\
\hline
$F_{\rho}/ m_{\rho}$ & first order perturb.  & 6\% \\
                     & zeroth order perturb.  & 7\% \\
                     & naive               & 10\% \\
\hline
$F_{\phi}/ m_{\rho}$ & first order perturb.  & 8\%  \\
                     & zeroth order perturb.  & 10\% \\
                     & naive               & 12\% \\
\hline
\end{tabular}
\caption{Estimated systematic errors in infinite volume quantities arising
from the correction from finite volume to infinite volume.}
\label{tab:syserrs}
\end{center}
\end{table}

\begin{table}
\begin{center}
\begin{tabular}{cllll}     \hline
 decay  & renorm. & finite volume     & infinite volume  & obs. \\ \hline
$f_{\pi} / m_{\rho}$ & first order perturb.  & $0.106 \pm 0.009$ &
                                               $0.106 \pm 0.014$ &  0.121 \\
                     & zeroth order perturb. & $0.110 \pm 0.009$ &         
                                               $0.110 \pm 0.015$ &  \\
                     & naive                 & $0.119 \pm 0.011$ & 
                                               $0.120 \pm 0.019$ &  \\ \hline
$f_K / m_{\rho}$     & first order perturb.  & $0.121 \pm 0.006$ & 
                                               $0.123 \pm 0.009$ & 0.148 \\
                     & zeroth order perturb. & $0.125 \pm 0.007$ &                 
                                               $0.127 \pm 0.010$ &  \\       
                     & naive                 & $0.141 \pm 0.008$ &                 
                                               $0.144 \pm 0.013$ &  \\ \hline
$F_{\rho}/ m_{\rho}$ & first order perturb.  & $0.177 \pm 0.021$ &
                                               $0.173 \pm 0.029$ & 0.199 \\ 
                     & zeroth order perturb. & $0.189 \pm 0.024$ &                 
                                               $0.184 \pm 0.036$ &   \\       
                     & naive                 & $0.191 \pm 0.030$ &                 
                                               $0.186 \pm 0.045$ &   \\ \hline
$F_{\phi}/ m_{\rho}$ & first order perturb.  & $0.217 \pm 0.019$ &  
                                               $0.253 \pm 0.035$ & 0.219 \\ 
                     & zeroth order perturb. & $0.234 \pm 0.023$ &                 
                                               $0.277 \pm 0.043$ &   \\       
                     & naive                 & $0.269 \pm 0.029$ &                 
                                               $0.328 \pm 0.052$ &   \\ \hline
\end{tabular}
\caption{Calculated values of meson decay constants
extrapolated to zero lattice spacing in finite volume,
then corrected to infinite volume, compared with observed
values.}
\label{tab:res}
\end{center}
\end{table}

\begin{table}
\begin{center}
\begin{tabular}{clll}     \hline
 decay  & finite volume     & infinite volume  & obs. \\ 
\hline
$f_{\pi} / f_K$ &  $0.875 \pm 0.034$ &
                   $0.864 \pm 0.065$ & 0.818 \\
$F_{\rho}/ f_K$ &  $1.467 \pm 0.170$ &
                   $1.412 \pm 0.229$ & 1.345 \\ 
$F_{\phi}/ f_K$ &  $1.795 \pm 0.131$ &  
                   $2.058 \pm 0.283$ & 1.480 \\ 
\hline
\end{tabular}
\caption{Calculated values of meson decay constants,
using first order perturbative renormalization,
extrapolated to zero lattice spacing in finite volume,
then corrected to infinite volume, compared with observed
values.}
\label{tab:resoverfk}
\end{center}
\end{table}

\clearpage

\end{document}